\documentclass{article}
\usepackage[utf8]{inputenc}
\usepackage{graphicx}
\usepackage{subcaption}
\usepackage{amsmath}
\usepackage{amssymb}
\usepackage{amsfonts}
\usepackage{textcomp}
\usepackage{url}
\usepackage{gensymb}
\usepackage{todonotes}
\usepackage{ctable}
\usepackage[switch, modulo]{lineno}
\usepackage{authblk}

\newcommand{\R}{\mathbb{R}}
\newcommand{\set}[1]{\mathcal{#1}}
\newcommand{\vect}[1]{\boldsymbol{#1}}

\title{Comparison of surface thermal patterns of horses and donkeys in IRT images}
\date{October, 2020}
\author[1,*]{Małgorzata Domino}
\author[2]{Michał Romaszewski}
\author[1]{Tomasz Jasiński}
\author[3]{Małgorzata Maśko}

\affil[1]{\small{Department of Large Animal Diseases and Clinic, Veterinary Research Centre and Center for Biomedical Research, Institute of Veterinary Medicine, Warsaw University of Life Sciences (WULS–SGGW), 02-787 Warsaw, Poland; malgorzata\_domino@sggw.edu.pl (M.D.); tomasz\_jasinski@sggw.edu.pl (T.J.);}}
\affil[2]{\small{Institute of Theoretical and Applied Informatics, Polish Academy of Sciences; 44-100 Gliwice, Poland; mromaszewski@iitis.pl (M.R.);}}
\affil[3]{\small{Department of Animal Breeding, Institute of Animal Science, Warsaw University of Life Sciences (WULS–SGGW), 02-787 Warsaw, Poland; malgorzata\_masko@sggw.edu.pl (M.M.)}}

\affil[*]{Correspondence: malgorzata\_domino@sggw.edu.pl; Tel.: +48-512-388-517}

\begin{document}
\maketitle

\begin{abstract}
Infrared thermography (IRT) is a valuable diagnostic tool in equine veterinary medicine however, little is known about its application in donkeys. The aim was to find patterns in thermal images of donkeys and horses, and determine if these patterns share similarities. The study was carried out on 18 donkeys and 16 horses. All equids underwent thermal imaging with an infrared camera and measuring the skin thickness and hair coat length. On the class maps of each thermal image, 15 regions of interest (ROIs) were annotated and then combined into 10 groups of ROIs (GORs). The existence of statistically significant differences between surface temperatures in GORs was tested both `globally' for all animals of a given species and `locally' for each animal. Two special cases of animals that differ from the rest were also discussed. 
Our results indicated that the majority of thermal patterns are similar for both species however, average surface temperatures in horses ($22.72\pm2.46\celsius)$) are higher than in donkeys ($18.88\pm2.30\celsius$). It may be related to differences in the skin and hair coat. We concluded, the patterns of both species are associated with GORs, rather than an individual ROI, with higher uniformity of donkeys patterns.    
\end{abstract}

\section*{Simple summary}
This study analyzes and compares the thermal patterns of horses and donkeys in infrared thermography (IRT) images. Thermal patterns were defined as statistically significant differences between groups of ROIs corresponding to areas with potential impact of large equine muscles. In our study, we used images of healthy and rested animals: sixteen horses and eighteen donkeys that formed a data set used in our experiments. We discussed our results,  compared thermal patterns between species and discussed special cases of animals identified as outliers. Our results support the thesis about similarities in the thermal patterns of horses and donkeys.

\section*{Keywords}
Infrared thermography; equids; thermal patterns; surface temperature; skin thickness; hair coat; 

\section{Introduction}

Infrared thermography (IRT) is a non-invasive imaging technique that allows to detect the radiant energy emitted by any object with a temperature above absolute zero. The radiated power detected by the thermal camera in the infrared spectrum is proportional to the fourth power of the object’s absolute temperature and is used to calculate the temperature of the target e.g. the surface of the animal's body. Infrared radiation is often presented as a thermogram which is an image where the color gradient corresponds with the distribution of surface temperatures ~\cite{soroko2018infrared}. Furthermore, the relationship of temperature gradients may create specific thermal patterns which may be used e.g. for assessing the influence of load on saddle fit in horses \cite{soroko2019evaluation} or the horses' response to the training \cite{masko2019pattern}.

IRT has been used as a diagnostic tool in equine veterinary medicine since the mid-1960s, particularly in the field of orthopedics, in the management of lameness~\cite{smith1964applications, eddy2001role, cetinkaya2012thermography, ciutacu2006igital}. The surface temperature changes, reflecting heat emitted from overloaded or injured tissue, were considered a valuable indicator for identifying areas of inflammation and blood flow alterations~\cite{kastberger2003infrared,  westermann2013effects}. This allows to detect temperature changes before they can be detected by palpation~\cite{alvarez2009back, bachi2018changes}, and before the onset of other clinical signs of injury \cite{bachi2018changes,seeherman1991use}. Therefore, in recent studies IRT was also applied to interpret changes in the surface temperatures of the thoracic region in the case of back pain diagnosis of equine athletes~\cite{fonseca2006thermography, soroko2012use}, as well as results of the impact of load on saddle \cite{soroko2019evaluation} or incorrect saddle fit~\cite{arruda2011thermographic}. Moreover, the usefulness of equine IRT in the assessment of transient stress response during training~\cite{becker2013cortisol,redaelli2019use} and sport competitions~\cite{valera2012changes, bartolome2013using} has been demonstrated. Equine IRT seems to be highly related to thermoregulation and the increase in blood flow due to exercise \cite{redaelli2019use}. During physical exercise, metabolic heat production increases as exercise intensity increases~\cite{hinchcliff2013equine}, and only a quarter of the energy used by a muscle is converted to mechanical energy. The remaining three quarters are dissipated as heat~\cite{simon2006influence}. Therefore, the radiant energy emitted from the horse's skin surface may be found as a product of basic metabolic processes, exercise, and pathological conditions. However, it should be kept in mind that the temperature measured from the body surface is related not only to the above internal conditions but also to the thermal properties of the skin and hair coat, and the thermal gradient between the skin surface and the environment~\cite{turner2001diagnostic, soroko2017effect}.

It is easy to see that IRT is widespread in the equestrian industry as a valuable tool to monitor the underlying circulation, tissue metabolism and local blood flow in response to different physiological, pathological, or environmental conditions. However, little or no attention has been paid to application of IRT to donkeys. The only work authors are aware of is the study of effect of season and age on daily rhythmicity of rectal temperature and body surface temperature during the cold-dry and hot-dry seasons in a tropical savannah~\cite{zakari2018daily}. For the infrared measurement, the infrared thermometer and seven landmarks adapted from equine IRT were used. Although this study evaluated differences in surface temperatures of donkeys of varying age groups under the changing environmental conditions, no studies to date have compared the thermal images of donkeys and horses obtained in the same circumstances. The scarcity of works on the imaging of donkeys motivates us to try to answer the question whether there are significant differences in thermal images of horses and donkeys. If the images of these animals were similar, it could suggest that intensively researched methods for analyzing equine images are applicable to donkeys.

We performed donkeys' and horses' imaging under the same environmental conditions. Following previous equine researchers, we evaluated body surface temperatures in healthy animals. The normal thermal image has already been described for e.g. the coronary band \cite{rosenmeier2012evaluation}, distal forelimbs joints \cite{machado2013standardization, soroko2017effect}, the thoracolumbar region \cite{tunley2004reliability}, and the back and pelvic regions \cite{pavelski2015infrared} in the horse. It showed a high degree of symmetry between the left and right sides of the body \cite{soroko2018infrared, soroko2017effect} and reproducibility over hourly, daily, and weekly intervals up to 90\% \cite{tunley2004reliability}. 

The thermal images were manually segmented into fifteen regions of interest (ROIs) corresponding to underlying large muscles. Since the phenomena observable in thermal images often include more than one ROI, we combined individual ROIs into groups of ROIs (GORs) and examined the differences in their mean temperatures. The differences, the occurrence of which has been statistically confirmed, constitute thermal patterns, which are the basis for the comparison of both species and analysis of special cases (outliers). This comparison is the main focus of our experiments. Our hypothesis is that thermal patterns of horses and donkeys are similar.

\section{Materials and Methods}
This section describes the methodology of acquiring,  describing and visualising our data, defining thermal patterns and assessing similarities between the patterns observed in thermal images of horses and donkeys. 

\subsection{Animals}
\label{sec:animals}
Eighteen donkeys (nine mares, seven geldings, and two stallions; mean age $7.78\pm3.04$ years; mean height $119.00\pm11.72$ cm) and sixteen horses/ponies called further horses (eight mares, six geldings, and two stallions; mean age $7.53\pm2.83$ years, mean height $137.40\pm9.33$ cm) participated in the study. The donkeys and horses are privately owned and are housed in the same stable located in southern Poland in Lubochów. The owners of the animals consent to carry out our research. The ethics approval was deemed unnecessary according to regulations of the II Local Ethical Committee on Animal Testing in Warsaw and the National Ethical Committees on Animal Testing because all procedures in the study were non-invasive and did not cause distress and pain equal to or greater than a needlestick. The equids were fed three times a day with a dose of hay personalized to each animal to maintain an optimal, healthy condition without obesity, and had daily access to a grassy paddock no shorter than 8 h per day. Both during the study and the month preceding the study, equids were not used in riding or harness. Before IRT imaging, physical examinations were conducted to ensure that the equids were free from a preexisting inflammatory condition. The imaging were carried out following the the international veterinary standards~\cite{purohit2009standards}. All donkeys and horses were clinically healthy, had no apparent back or lameness problems, and demonstrated conformation typical of their species and comparable growth. Two donkeys were excluded at the stage of preliminary examination. The first of them due to plaque losses in the hair coat caused by abrasions in transport in the week preceding the study, wheres the second due to the significantly longer hair length ($7.6\pm1.2$ cm) in comparison to the the hair length of the other donkeys ($3.4\pm0.7$ cm). Finally, sixteen donkeys were qualified for the formal analysis, however, an analysis of the two donkeys deviating from the accepted uniform appearance is included in the discussion section.

\subsection{Data collection}
To ensure the best possible conditions for comparison of collected thermal images, the skin thickness, the hair coat length, and the constant thermal gradient between the skin surface and the environment were taken into account. The study was performed in middle September, and all measurements were taken on the same day under the same circumstances (ambient temperature 20.2\celsius; humidity 45\%). 
A total of 68 images were taken in a closed space, protected from wind and sun radiation, to minimize the influence of external environmental conditions~\cite{satchell2015effects}. The imaged area was brushed and dirt and mud were removed 15 minute before imaging. The thermal images were acquired on the left and right sides at a 90\degree camera angle from a distance of approximately 2 m from the donkey or the horse. During each imaging session two images of each individual were taken. The images were positioned on the centre of the trunk. Images were taken using an infrared radiation camera (FLIR Therma CAM E25, Brazil) with an emissivity (e)~0.99, by the same researcher (MM). The temperature range was standardized in the professional software (FLIR Tools Professional, Brazil) during the pre-processing of images at $10$-$30\celsius$ level. 

\begin{figure}
	\centering
	\begin{subfigure}[b]{0.49\textwidth}
	\includegraphics[width=1.0\linewidth]{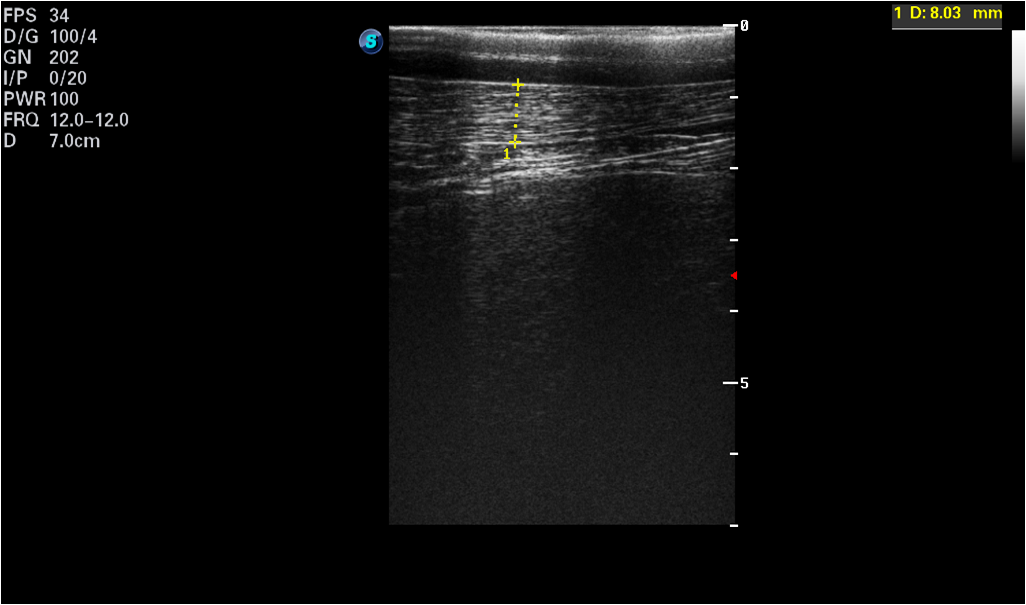}
	\caption{SF-Skin of horse}
	\end{subfigure}
	\begin{subfigure}[b]{0.49\textwidth}
	\includegraphics[width=1.0\linewidth]{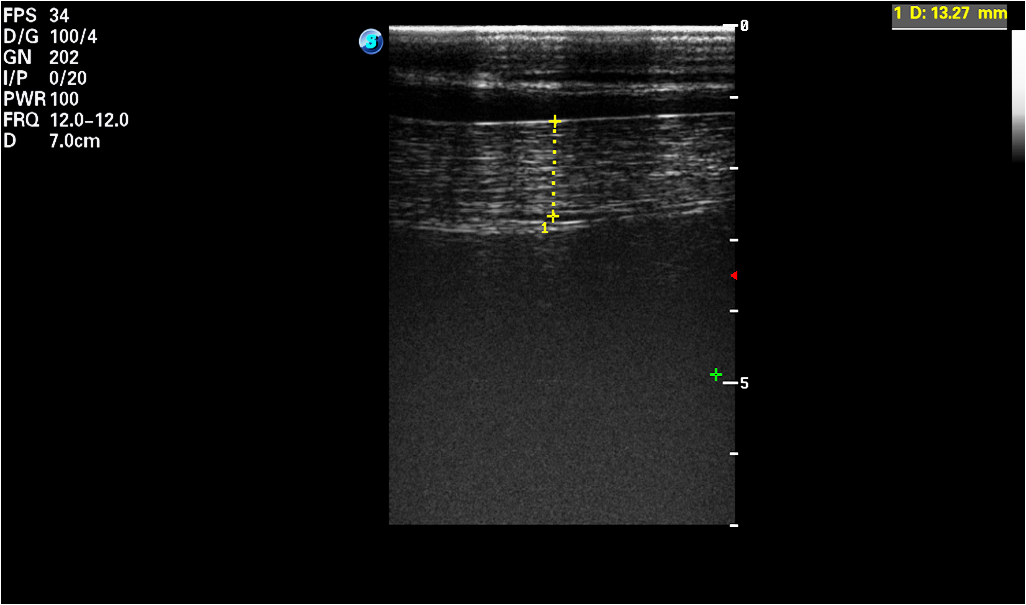}
	\caption{SF-Skin of donkey}
	\end{subfigure}		
	\caption{Example of an ultrasonographic image taken over the 3rd lumbar vertebra: (a) the horse \emph{H.1}; (b) the donkey \emph{D.3}. A subcutaneous fat plus skin thickness (SF-Skin) is highlighted.}
	\label{fig:usg}
\end{figure}
\ctable[
cap     = Animal features,
caption = Measured features (mean$\pm$SD) of horses (H.1-H.16) and donkeys (D.1-D.16): the length of hair coat and the thickness of the subcutaneous fat plus skin (SF-Skin).,
label   = tab:animalsft,
pos     = h]
{lcc}
{\tnote[]{Different superscript letters indicate significant differences between Horses and Donkeys for Hair coat (a, b) and SF-Skin (c, d) respectively according to the Mann-Whitney-Wilcoxon (MWW) test}}{\FL
Animals&Hair coat [cm]&SF-Skin [mm]\ML
Donkeys&$3.39\pm0.46$\tmark[a]&$12.01\pm0.83$\tmark[c]\NN
Horses&$1.78\pm0.38$\tmark[b]&$8.80\pm0.87$\tmark[d]\ML
p-value&$<0.0001$&$<0.0001$\LL
}
After each IRT imaging, the ultrasonographic image was taken with an ultrasound scanner (SonoScape S9, SonoScape, Shenzhen, China) using a linear 5-12 MHz transducer (L752, onoScape, Shenzhen, China). Ultrasound scans were performed with the transducer placed at the animal´s back, over the 3rd lumbar vertebra, perpendicular to the backbone; all the images were collected on the left side of the animal~\cite{silva2016relationships}. The hair was trimmed at the measurement place and ultrasound gel (Aquasonic 100, Parker, USA) was used as a coupling medium. The real time ultrasonographic examination was freezed, the image was saved, and the subcutaneous fat (SF) plus skin thickness (SF-Skin) measurement were obtained. An example of an ultrasonographic image is presented in~Fig.\ref{fig:usg}. The hair coat samples were taken from the midneck about 5 cm below the base of the mane. The length of individual hairs were determined from a random sample of five pulled strands, including the roots \cite{osthaus2018hair}.
Average hair coat length and SF-Skin values are presented in Tab.~\ref{tab:animalsft}.

\subsubsection{Dataset preparation}
\label{sec:rois}
\begin{figure}
	\centering
	\begin{subfigure}[b]{0.32\textwidth}
	\includegraphics[width=1.0\linewidth]{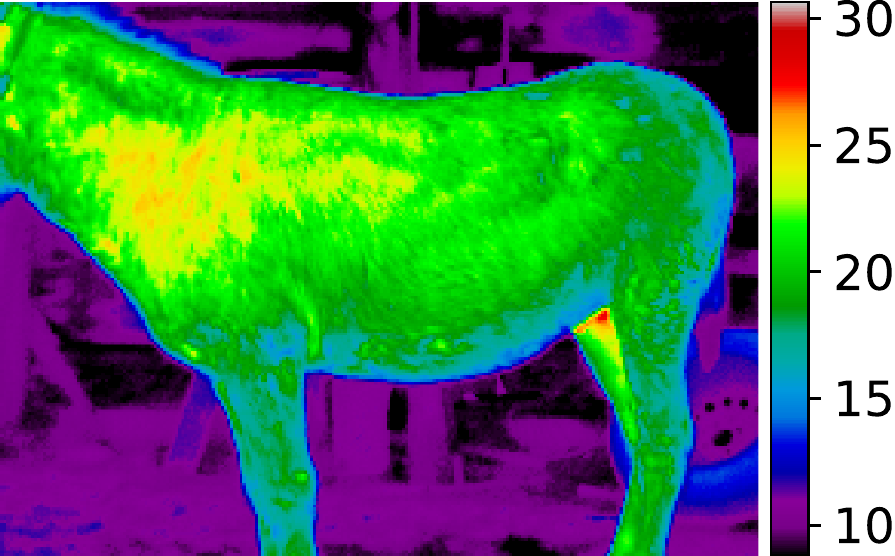}
	\caption{Thermal map}
	\end{subfigure}
	\begin{subfigure}[b]{0.27\textwidth}
	\includegraphics[width=1.0\linewidth]{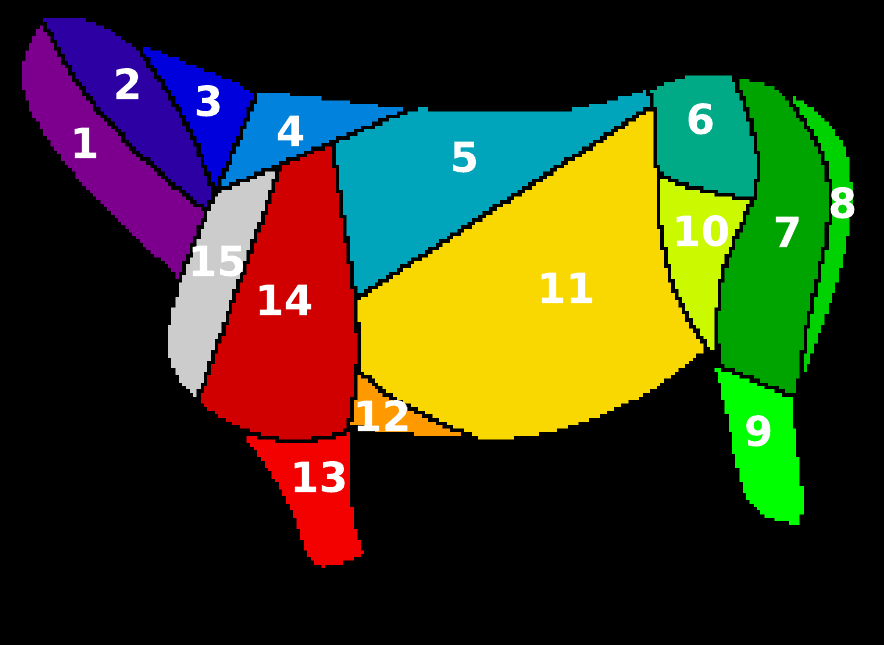}
	\caption{ROIs}
	\end{subfigure}
	\begin{subfigure}[b]{0.32\textwidth}
	\includegraphics[width=1.0\linewidth]{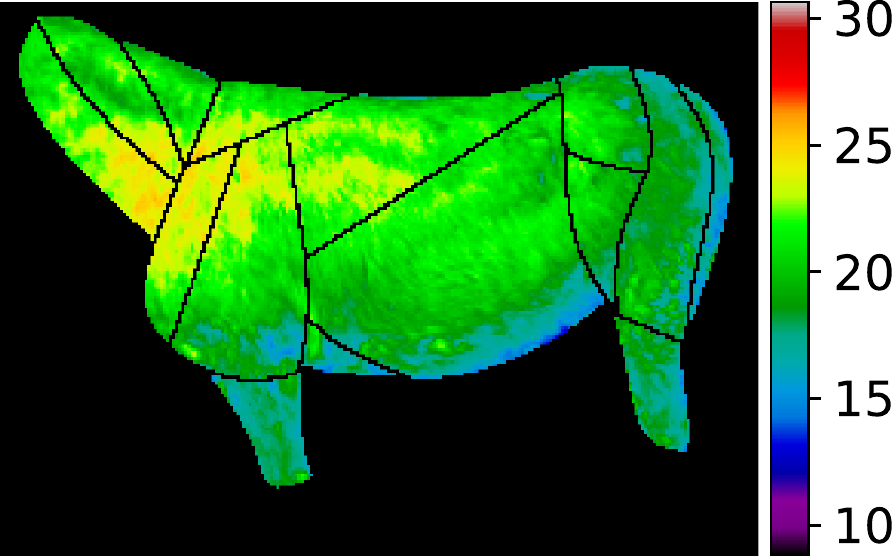}
	\caption{Extracted pixels}
	\end{subfigure}	
	\caption{Visualisation of a donkey \emph{D.3}: (a) thermal data from the camera as a thermal map; (b) annotated classes corresponding to selected ROIs (see Sec.~\ref{sec:rois}); (c) extracted areas, used in our experiments.}
	\label{fig:demo}
\end{figure}

Based on collected data, a data set was prepared that was later used in our experiments. The data set consists of images from a thermal camera and the corresponding annotations in the form of class maps of main muscle areas. Every thermal image is a table of $320x240$ pixels. The value in each pixel corresponds to the measured temperature value. A corresponding class map is a table, where the value in every pixel is the ROI number (or zero areas without annotations). An example class map is presented in~Fig.~\ref{fig:demo}. The class maps were produced by hand annotating the fifteen identified regions of interest (ROIs) in each image. The following ROIs corresponding to the underlying large muscles were annotated:

\begin{enumerate}
    \item ROI~1 \emph{m. brachiocephalicus} - a parallelogram-shaped area from the lateral surface of the Atlas, behind the angle of the mandible, to the regio supraspinata of the scapula.
    \item ROI~2 \emph{mm. splenius capitis and cervicis} - a triangle-shaped area from the lateral surface of the axis to the regio supraspinata of the scapula above ROI 3.
	\item ROI~3 \emph{m. trapezius pars cervicalis} - a triangle ranged from the middle of the neck to the regio cartilaginis of the scapula and along the regio supraspinata of the scapula up to two-thirds of the length of the scapula.
	\item ROI~4 \emph{m. trapezius pars thoracica} - a triangle ranged from the the regio cartilaginis of the scapula along the regio supraspinata of the scapula up to one-thirds of the length of the scapula.
	\item ROI~5 \emph{m. latissimus dorsi} - a triangle-shaped area from the regio infraspinata of the scapula, up to two-thirds of the length of the scapula, along the back to the tuber coxae.
	\item ROI~6 \emph{mm. glutei} (superficialis and medius) - an  irregular area in the regio tuberis coxae.
	\item ROI~7 \emph{m. biceps femoris} - an oblong s-shaped area in the regio femoris cranially from the m. semitendinosus.
	\item ROI~8 \emph{m. semitendinosus} - an oblong s-shaped area in the regio femoris caudally from the m. biceps femoris.
	\item ROI~9 \emph{mm. in regio cruris} - a rectangular-shaped area in the regio cruris between articulatio genus and articulatio tarsi.
	\item ROI~10 \emph{m. tensor fasciae latae} - an  irregular area between the in the regio tuberis coxae and the flank.
	\item ROI~11 \emph{m. obliquus externus abdominis} - a trapezoid-shaped area from the lower two-thirds of the regio infraspinata of the scapula to the tuber coxae and the regio of processus xiphoideus sterni.
	\item ROI~12 \emph{m. pectoralis transversus} - a triangle-shaped area behind the regio of olecranon to the regio of processus xiphoideus sterni.
	\item ROI~13 \emph{mm. in regio antebrachi} - a rectangular-shaped area in the regio antebrachi between articulatio humeri and articulatio cubiti.
	\item ROI~14 \emph{m. pectoralis descendens} - an irregular area in the projection of regio infraspinata of the scapula.
	\item ROI~15 \emph{m. deltoideus} - an irregular area in the projection of the regio supraspinata of the scapula.   
\end{enumerate}

\subsubsection{Dataset availability}
In order to facilitate replication of our results the data set\footnote{Data set location: \url{10.5281/zenodo.4085075}} and the experimental source code\footnote{Source code location: \url{https://github.com/iitis/thermal_patterns.git}} have been made available to the public under an open license.

\subsection{Thermal patterns in IRT images of horses and donkeys}
Our main goal was to find patterns in thermal images of both species and determine if these patterns share similarities. We define a thermal pattern as a statistically significant difference between two areas composed of ROI groups. Here we describe our methodology of finding and confirming the relevance of thermal patterns. 

\subsubsection{Testing statistical significance of temperature differences}
\label{sec:statistical_significance}
In this work our claims usually involve comparison of temperatures between areas (subsets of pixels) in a thermal image or images e.g. when we state `animals A are warmer than animals B in area C' this means that based on our sample, the surface temperature of animals A is on average higher in this region. Therefore, when we claim that such statement is statistically significant, to confirm this we use the Mann–Whitney–Wilcoxon (MWW) test~\cite{derrick2017comparing} (one-sided) which is a non-parametric statistical hypothesis test that allows to compare two related sequences of samples. We have chosen a non-parametric test due to the fact that temperature distributions in our ROIs are diverse and often non-Gaussian. The two sequences are obtained by randomly, uniformly sampling the compared ROI or groups of ROIs. The number of samples is the size of the smaller set -- when comparing individual ROIs between animals, the difference in sample count is usually less than 10\%. Unless stated otherwise, we require that the p-value $p<0.001$.

\subsubsection{Finding thermal patterns in animal species}
\label{sec:thermal_patterns_method}
The general idea is, that while we expect global differences between animal species e.g. on average, one species can have a higher surface temperature the other, we are interested in the existence of repeated dependencies between surface temperatures in different body areas of a given species. Such dependencies can form a pattern that may be compared for both species. To identify these patterns, we use the following methodology:

\paragraph{Combining ROIs into groups of ROIs (GORs)}
\begin{figure}
	\centering
	\begin{subfigure}[b]{0.24\textwidth}
	\includegraphics[width=1.0\linewidth]{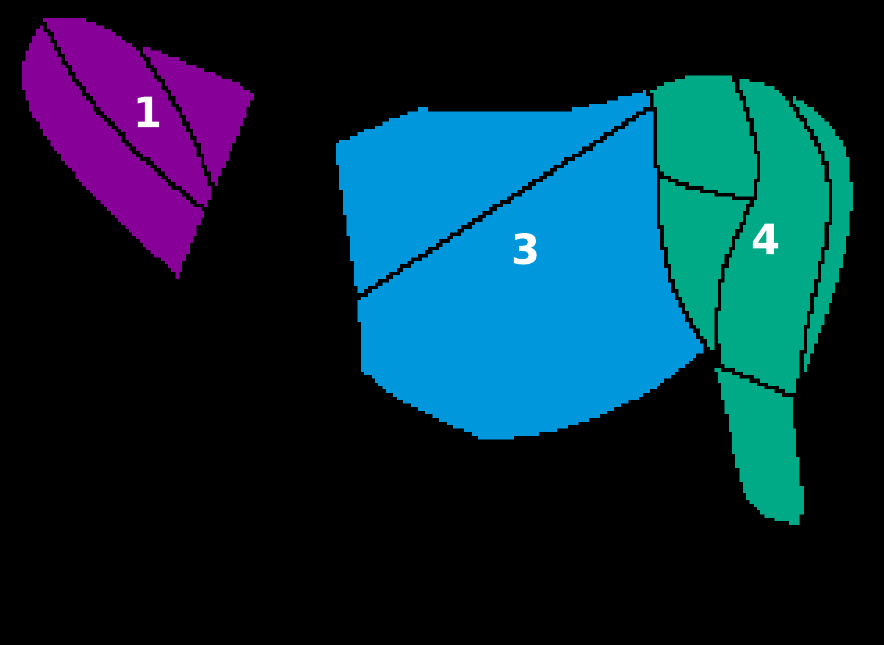}
	\end{subfigure}
	\begin{subfigure}[b]{0.24\textwidth}
	\includegraphics[width=1.0\linewidth]{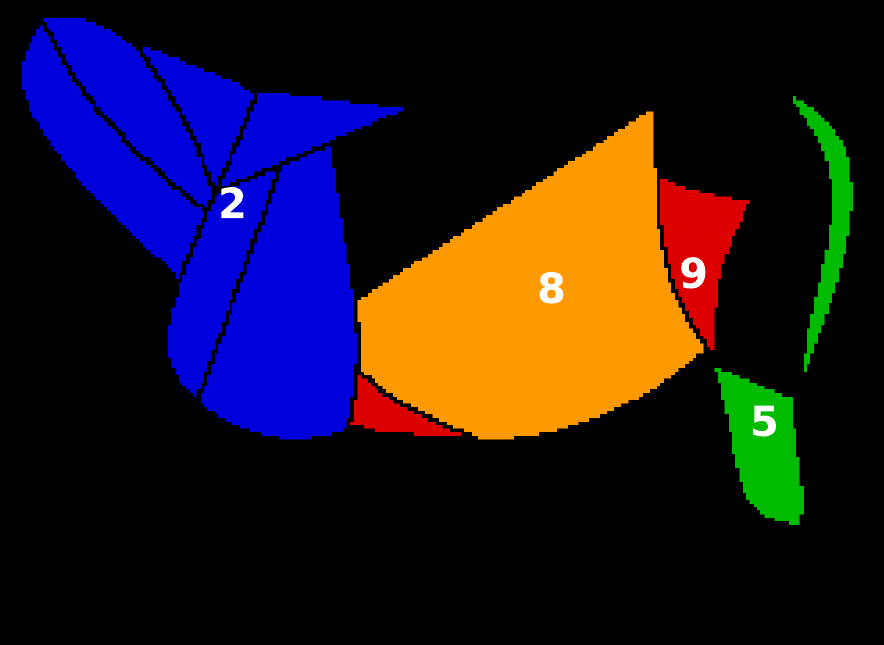}
	\end{subfigure}
	\begin{subfigure}[b]{0.24\textwidth}
	\includegraphics[width=1.0\linewidth]{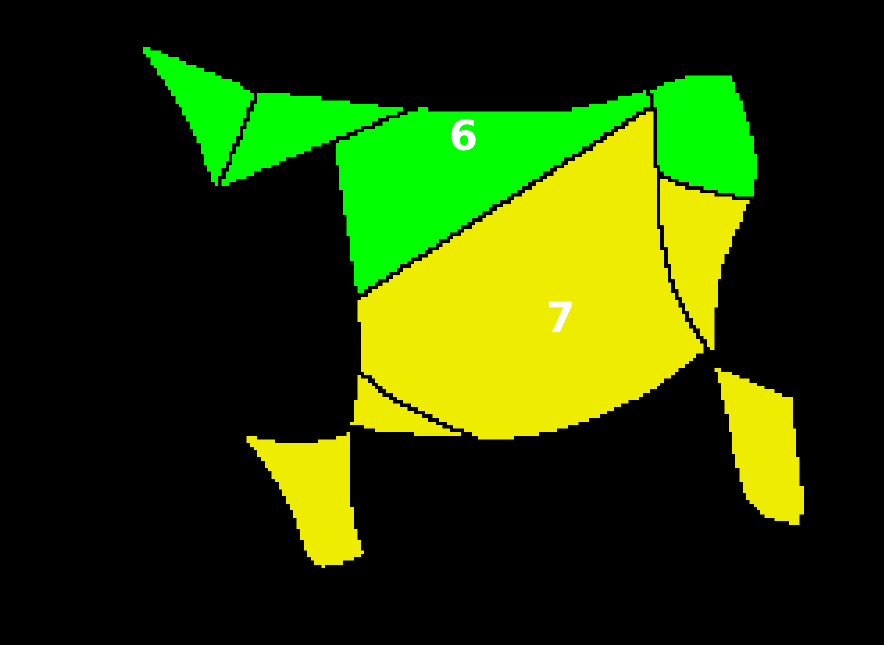}
	\end{subfigure}
    \begin{subfigure}[b]{0.24\textwidth}
	\includegraphics[width=1.0\linewidth]{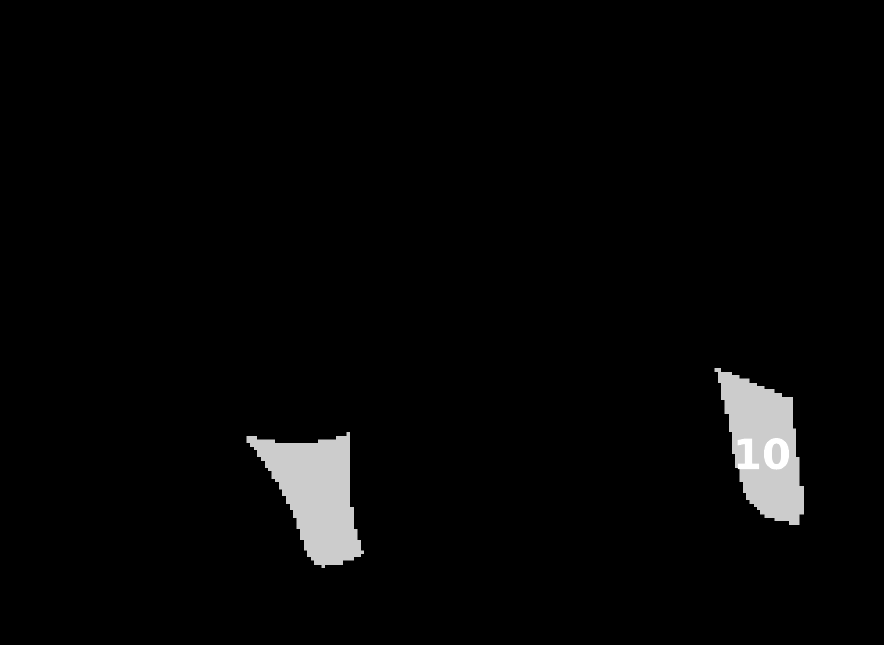}
	\end{subfigure}		
	\caption{Visualisation of a donkey \emph{D.3}, divided into Groups of ROIs (GORs).}
	\label{fig:demo_GORs}
\end{figure}

In the first step, based on our observation that visible patterns in thermal images from our data set are often located in several ROIs, we have designated manually 10 groups of ROIs (GORs) for our analysis. The designed GORs, presented in Fig.~\ref{fig:demo_GORs} were as follows: 

\begin{enumerate}
\item GOR~1 \emph{Neck}, ROIs $\{1,2,3\}$ -- represented an area with an impact of muscles located cranially from the cranial border of scapulae
\item GOR~2 \emph{Frontquarter}, ROIs $\{1,2,3,4,14,15\}$ -- represented an area with an impact of muscles located cranially from the spinous processes of scapulae
\item GOR~3 \emph{Trunk}, ROIs $\{5,11\}$ -- represented an area with an impact of muscles layed between the caudal border of scapulae and the vertical line defined by tuber coxae, expecting area of \emph{m. pectoralis transversus} 
\item GOR~4 \emph{Hindquarter}, ROIs $\{6, 7, 8, 9, 10\}$ -- represented an area with an impact of the examined muscles of the pelvic limbs layed caudally from the vertical line defined by tuber coxae
\item GOR~5 \emph{Rump}, ROIs $\{8, 9\}$ -- represented an area with an impact of two ROIs of the pelvic limbs layed the most caudally
\item GOR~6 \emph{Dorsal aspect}, ROIs $\{3, 4, 5, 6\}$ -- collected the area with impact of muscles located above the horizontal line halfway up the trunk
\item GOR~7 \emph{Ventral aspect}, ROIs $\{9, 10, 11, 12, 13\}$ -- collected the area with impact of muscles located below the horizontal line halfway up the trunk
\item GOR~8 \emph{Abdomen}, ROI $\{11\}$ -- represented an area with an impact of muscles layed between the caudal border of scapulae and the vertical line defined by tuber coxae, expecting area of \emph{m. pectoralis transversus} and \emph{m. latissimus dorsi}
\item GOR~9 \emph{Groins} (Girth and Flank), ROIs $\{10, 12\}$ -- represented two areas most covered by large muscles of thoracic and pelvic limbs with so girth area and flank area
\item GOR~10 \emph{Legs}, ROIs $\{9, 13\}$ -- represented two areas with an impact of muscles of the proximal parts of limbs, both thoracic and pelvic
\end{enumerate}

\paragraph{Comparing GORs temperatures}
Our goal is to compare average temperatures between designated groups of ROIs and test whether the difference is statistically significant. We do it as follows: 

We have a set of animals of a given species $\set{A}=\{a_1,..,a_{16}\}$ and a set of groups of ROIs defined in the previous paragraph $\set{G}=\{g_1,..,g_{10}\}$. 
A set $\set{T}^a_g$ is a set 
of pixels (temperatures) of an animal $a\in\set{A}$ from a group $g\in\set{G}$ and the mean value of pixels in a set is denoted by $\delta$ e.g. $\delta(\set{T}^a_g)$. For every pair of groups $(i,j)\in\set{G}\times\set{G}$ we compute a difference in average values of temperatures in those groups for all animals i.e. $$\Delta_{(i,j)}=\delta\left(\bigcup\limits_{k\in\set{A}}\set{T}^j_k \right)-\delta\left(\bigcup\limits_{k\in\set{A}}\set{T}^j_k \right).$$ These differences are presented in our results as a matrix of differences $\vect{M}_\Delta\in\R^{|\set{G}|\times|\set{G}|}$. Note that the matrix $\vect{M}_\Delta$ is not symmetric as it shows temperature differences and not their absolute values.

Our next step is to apply the MWW test, described in Sec.~\ref{sec:statistical_significance} to verify the statistical significance of the difference $\Delta_{(i,j)}$ for every pair of groups $(i,j)\in\set{G}\times\set{G}$. We do it in two ways:
\begin{enumerate}
    \item Globally - for every pair $(i,j)\in\set{G}\times\set{G}$ we apply the MWW test to the whole population i.e. we compare the union of sets $\bigcup\limits_{k\in\set{A}}\set{T}^k_i$ with the union of sets $\bigcup\limits_{k\in\set{A}}\set{T}^k_j$.
    \item Locally - for every pair $(i,j)\in\set{G}\times\set{G}$ we apply the MWW test separately for every animal $a\in\set{A}$, by comparing the set $\set{T}^a_i$ with the set $\set{T}^a_j$. For our data set this results in $16$ tests for every pair.
\end{enumerate}

In our results, outcomes of the MWW tests supplement the presented matrix of differences: results of the `global' test are presented in the matrix itself (with a bold font) while results of `local' tests are presented as a separate matrix $\vect{M}_L\in\R_{+}^{|\set{G}|\times|\set{G}|}$, where the value in every cell represents the number of animals for which the difference was significant.

\paragraph{Thermal patterns}
For a given animal species we treat a statistically significant difference $\Delta_{(i,j)}$ between two groups of ROIs $(i,j)\in\set{G}\times\set{G}$ as a thermal pattern. A thermal pattern can thus be interpreted as a statement that based on our data e.g.the \emph{Rump} area is colder than the \emph{Neck} area. If the significance is confirmed by the `global' test but not for every animal in the `local' test (i.e. the value in the matrix $\vect{M}_L$ for this pair is not  $16$), it means that while the pattern emerges in a population, it is susceptible to individual differences of animals, and that there are animals that do not show it. If the pattern also appears individually in most (or all) of the animals tested, we consider it to be more stable and reliable.

Analysis of the structure of matrices $M_\Delta$ and $M_L$ for both species will be the basis of our discussion about similarities in their IRT images.

\subsection{Visualisation techniques}
Here we describe visualisation techniques used to illustrate our observations.

\subsubsection{Thermal images visualisation}
In order to visualize the visible structures in IRT images from our dataset, the temperature was presented in the form of a color map, modeled on the visible part of the electromagnetic spectrum, i.e. going from violet to red. To improve the clarity of images, the zero values representing the areas outside the ROIs are shown in black.
By manipulating the color map threshold values (assigned to its extreme colors), we visualise patterns common to all  animals or highlight patterns specific to a particular animal. An example of such a visualization is presented in~Fig.~\ref{fig:heatmap_relative}.

Temperature distributions within a specific ROI are visualized using histograms where the y-axis is presented as a probability density, i.e. bin counts are divided by a total number of counts. Alternatively, we use boxplots where the box extends from the lower to the upper quantile values. The line in the boxplot denotes the median, the whiskers denote the range of~$\lbrace q_1-1.5*(q_3-q_1),q_3-1.5*(q_3-q_1)\rbrace$ where $q_1$ $q_3$ denote the first and the third quartiles and circles denote outliers.

\subsubsection{Data visualisation}
An individual animal in our data set can be represented by a vector $\vect{v_i\in\R^d}$ of $d$ features corresponding e.g. to means or variances of temperatures in every ROI which leads to $d\geq15$. Extraction and visualisation of data structures in high-dimensional space is often performed by using the Principal Component Analysis~\cite{hotelling1933analysis} (PCA) and projecting data onto the first principal components. However, PCA uses the covariance matrix of data and since our data set contains limited number of examples, this makes computation of a reliable covariance matrix difficult. Therefore, to present our data we use the t-Distributed Stochastic Neighbor Embedding (t-SNE)~\cite{maaten2008visualizing} algorithm that visualises data by giving each example a location in a two-dimensional map. An important feature of the t-SNE is that its output is non-deterministic which results from the fact that the optimisation problem solved by the technique has a cost function that is not convex. Since, in this work t-SNE is only used to highlight patterns emerging in data, we consider this acceptable and presented visualisations are selected as representative examples after several executions of t-SNE.

We will analyse the structure of t-SNE visualisation to investigate whether ROI features are characteristic for both species and allow to distinguish between them. Features will be extracted with common statistics such as the mean, standard deviation, kurtosis and skewness. In addition, we will also check the impact of mean normalisation by removing the global temperature of an animal from all pixels. This last experiment is aimed to test the distinctiveness of differences between ROI temperatures of one animal.

\section{Results}
This section describes our experiments and presents their results.

\subsection{Experiments}

Unless stated otherwise, in all our thermal map visualisations, e.g. in the upper row of the Fig.~\ref{fig:heatmap_relative}, presented color map temperature values $t_c$ were limited either to the common range of  $t_c\in\langle8.8,30.65\rangle$\celsius, which are extreme values in annotated ROIs for all animals included in the study\footnote{Temperatures in ROIs -- horses: $t_h\in\langle10.64,30.65\rangle\celsius, E(t_h)=22.72\pm2.46\celsius$,\\ donkeys: $t_d\in\langle8.8,29.56\rangle\celsius, E(t_h)=18.88\pm2.30\celsius$} (not counting animals \emph{D.17}, \emph{D.18}, for reasons explained in Sec.~\ref{sec:animals}). 

Alternatively, they were selected as extreme temperature values in ROIs for a given animal to highlight features of visible thermal patterns - these special cases are clearly indicated. When applying t-SNE for data visualisation, its perplexity parameter was set to the value of 5.

\subsubsection{Implementation}
Experiments were implemented in Python 3.6.9 using libraries: numpy 1.16.4, scipy 1.3.1, scikit-learn 0.22.1, matplotlib 3.2.2

Experiments were conducted using a computer with\\ Intel(R)~Core~i7-5820K~CPU~@~330GHz with 64GB of RAM and with the Windows 10 Pro system. The running time of experiments could be measured in seconds.

\subsection{Experimental results}
\begin{figure}
	\centering
	\begin{subfigure}[b]{0.49\textwidth}
    \includegraphics[width=1.0\linewidth]{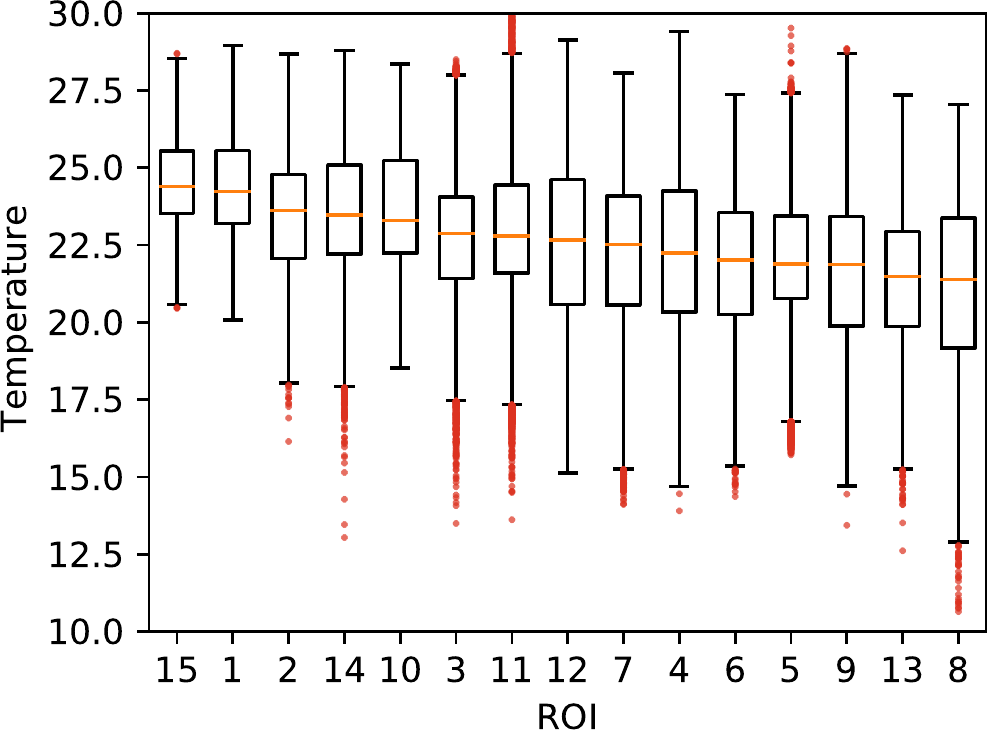}
    \caption{Horses}
    \end{subfigure}
    \begin{subfigure}[b]{0.49\textwidth}
    \includegraphics[width=1.0\linewidth]{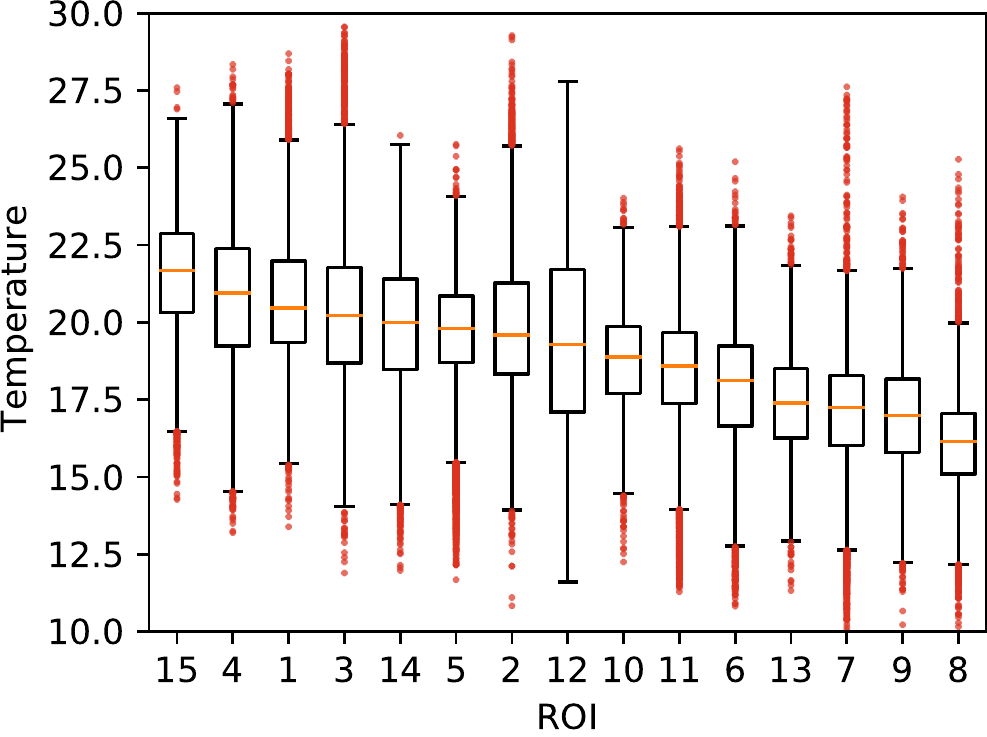}
    \caption{Donkeys}
    \end{subfigure}
	\caption{Visualisation of temperatures in ROIs for of all animals: (a) horses; (b) donkeys. ROIs were ordered by their medians.}
	\label{fig:box_rois}
\end{figure}
\begin{figure}
	\centering
    \begin{subfigure}[b]{0.49\textwidth}
    \includegraphics[width=1.0\linewidth]{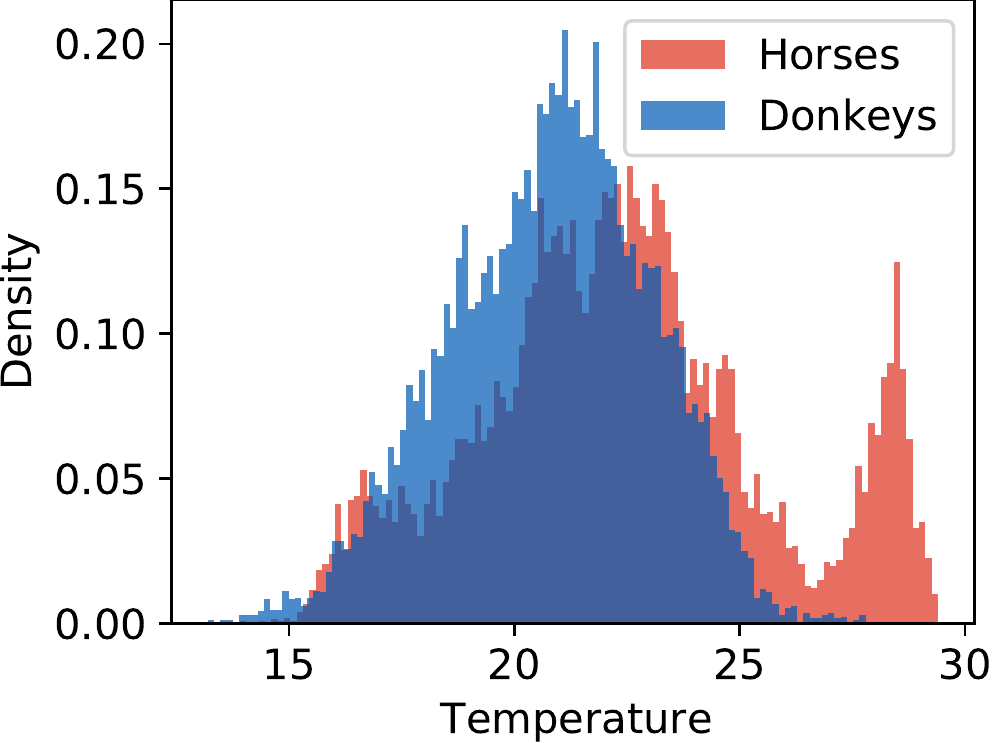}
    \caption{ROI 4}
    \end{subfigure}
    \begin{subfigure}[b]{0.49\textwidth}
    \includegraphics[width=1.0\linewidth]{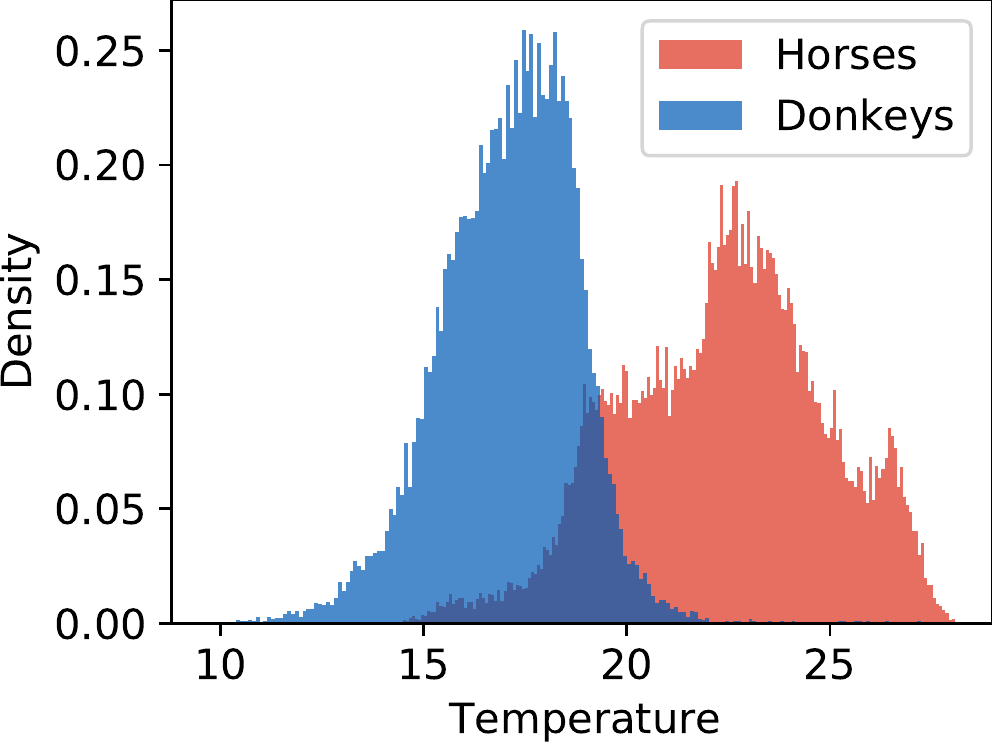}
    \caption{ROI 7}
    \end{subfigure}
	\caption{Histograms of temperatures for two ROIs where the difference $\Delta_{t}$ between mean values for the two animal species is: (a) the smallest (ROI 4, $\Delta_{t}=1.59$) and (b) the largest (ROI 7, $\Delta_{t}=5.26$).}
	\label{fig:histo_min_max}
\end{figure}
A comparison of temperatures between ROIs is presented in Fig.~\ref{fig:box_rois}. The immediate observation is that surface temperatures for horses are, on average, higher than for donkeys, which is confirmed as statistically significant for every ROI (MWW test, see. Sec.~\ref{sec:statistical_significance}). We can see that there are considerable variances in ROIs temperatures, and many outliers. Fig.~\ref{fig:histo_min_max} presents example histograms for two ROIs where differences in mean temperatures of horses and donkeys are most extreme. We can see how temperature distributions can be multi-modal which results from individual differences in animal surface temperatures. 
Histograms of temperatures for all ROIs can be found in Fig.~\ref{fig:histo_all} in the Appendix. 

\begin{figure}
	\centering
	\begin{subfigure}[b]{0.49\textwidth}
	\includegraphics[width=1.0\linewidth]{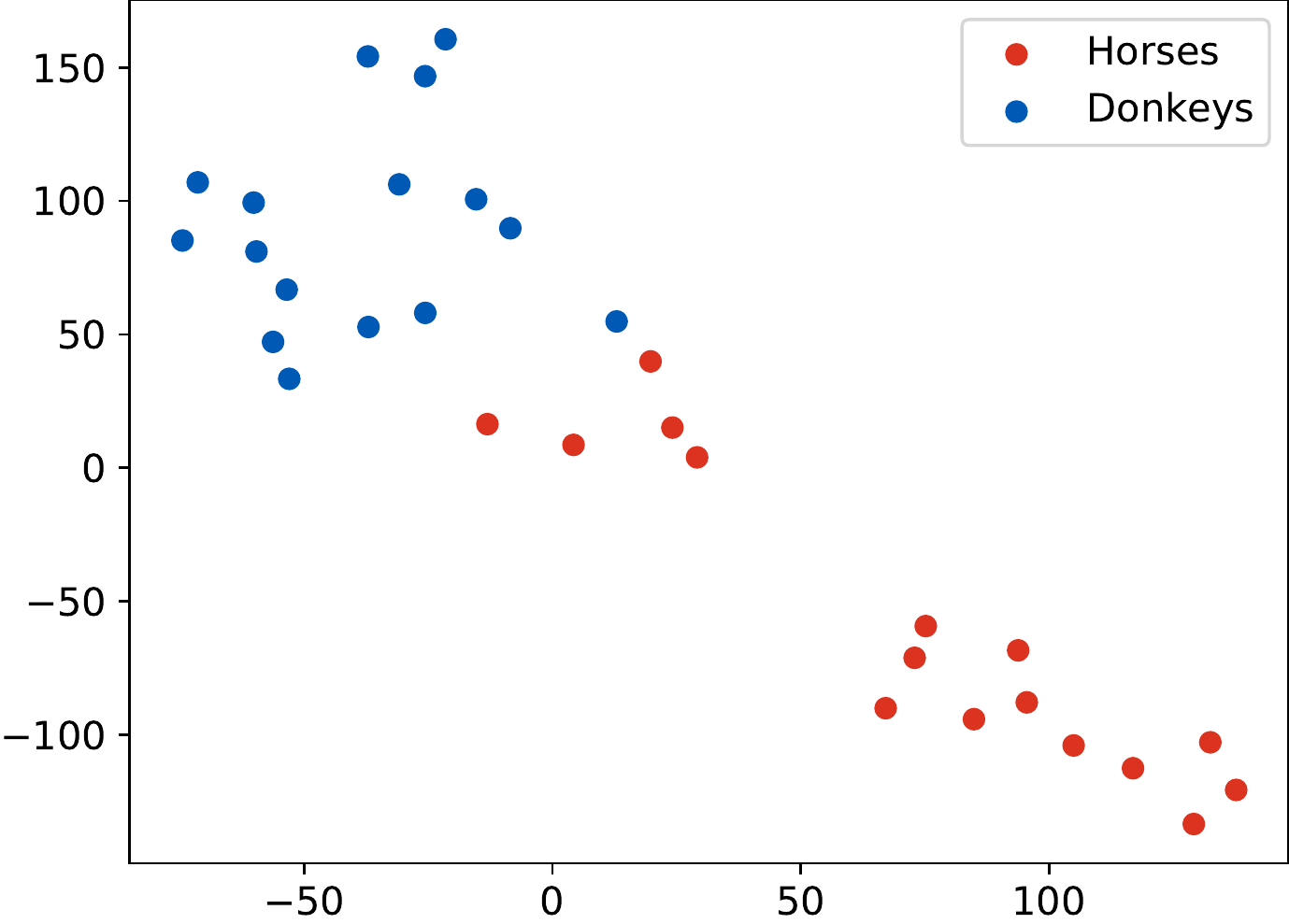}
	\caption{Mean} 
	\end{subfigure}
	\begin{subfigure}[b]{0.49\textwidth}
	\includegraphics[width=1.0\linewidth]{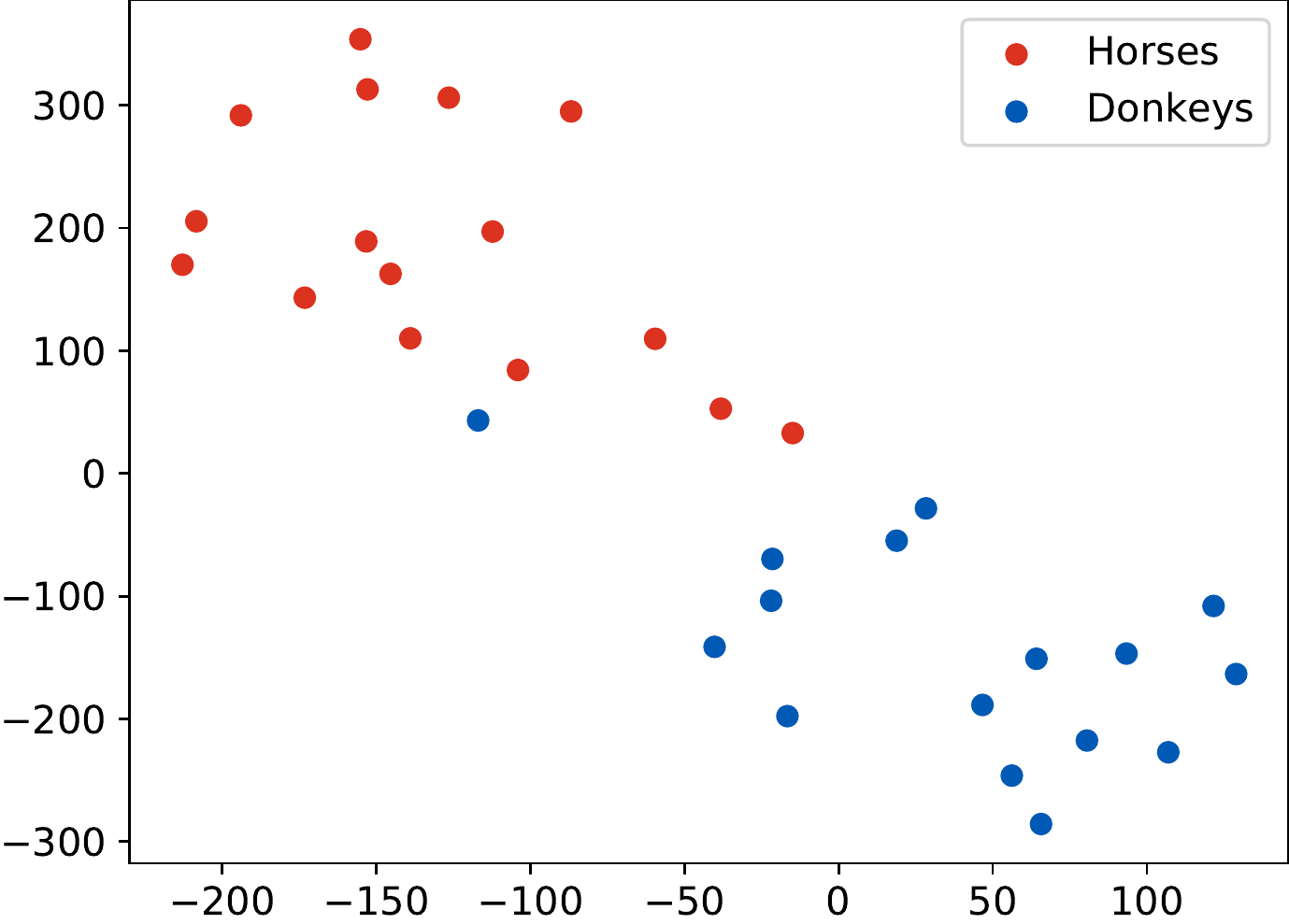}
	\caption{Std} 
	\end{subfigure}
	\begin{subfigure}[b]{0.49\textwidth}
	\includegraphics[width=1.0\linewidth]{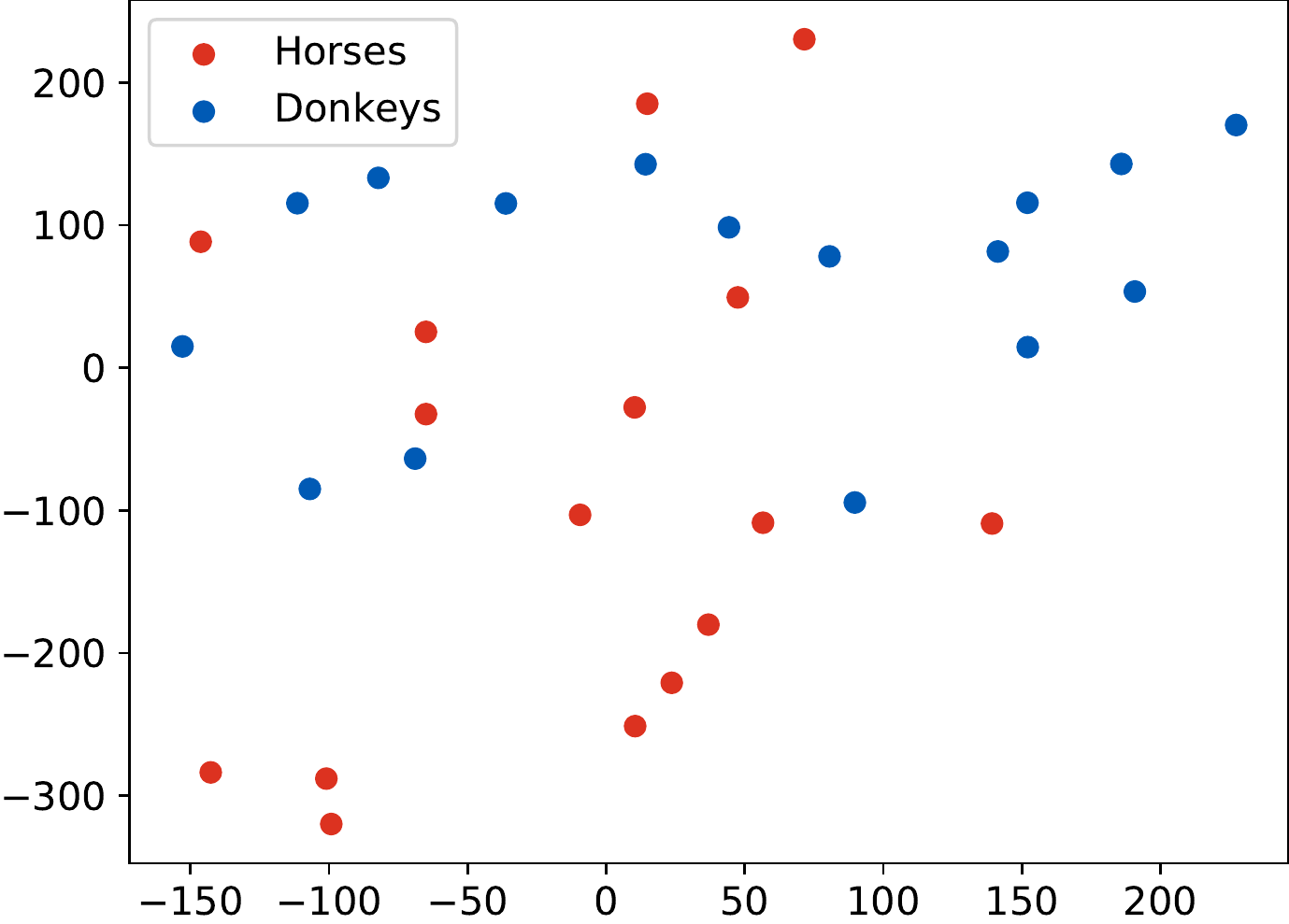}
	\caption{Kurtosis} 
	\end{subfigure}
	\begin{subfigure}[b]{0.49\textwidth}
	\includegraphics[width=1.0\linewidth]{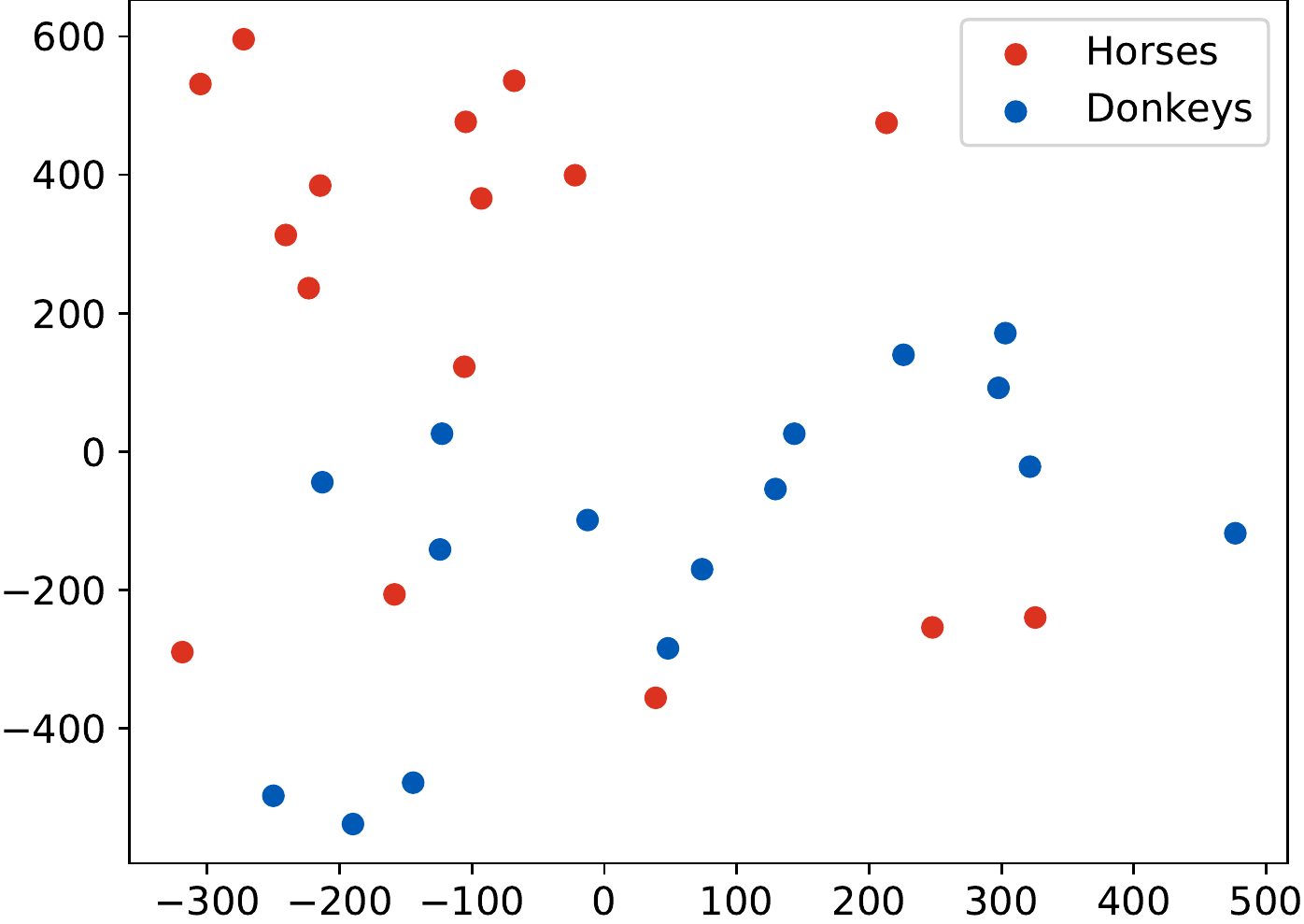}
	\caption{Mean (normalised)} 
	\end{subfigure}
	\caption{T-SNR visualisation of the data set. Every dot represents an animal described with features extracted from pixels of its 15 ROIs. Plots present different feature extraction statistics: (a) the mean; (b) the standard deviation; (c) the kurtosis; (d) the mean, after removing the global mean temperature of an animal from all pixel values. Notice that features in plots~(a) and~(b) seem more distinctive for both species which results in more apparent clusters. However, the classes of examples in these clusters are mixed.}
	\label{fig:tsnr}
\end{figure}

In order to visualise ROI features and assess their potential for distinguishing between species we used the t-SNR visualisation. Example results are presented in Fig.~\ref{fig:tsnr}. We can see that for the mean or standard deviation of temperatures in ROIs, examples form two clusters corresponding to the species of animals. However, some examples are in the wrong cluster, which suggests that a subset of animals could be missclassified. For features based on skewness and kurtosis no consistent structures are observed, which indicates that these features by themselves may not be a good basis for classification. For the normalized temperature, while grouping of examples in classes could be observed, a subset of examples is usually far from their clusters. These results indicate that simple features of ROIs may not be sufficient to accurately classify animal species and that a combination of features and/or more complex statistics may be needed for this task.

\begin{figure}
	\centering
	\includegraphics[width=0.24\linewidth]{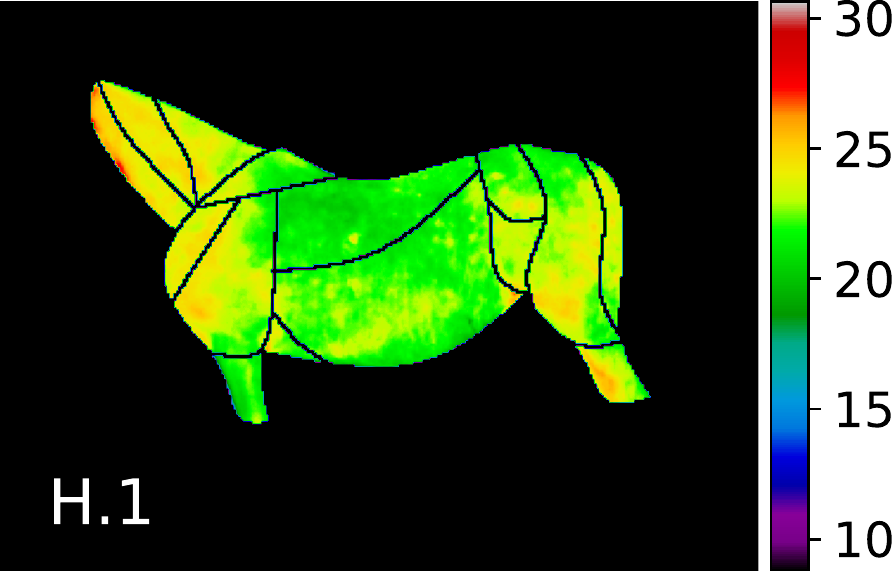}
	\includegraphics[width=0.24\linewidth]{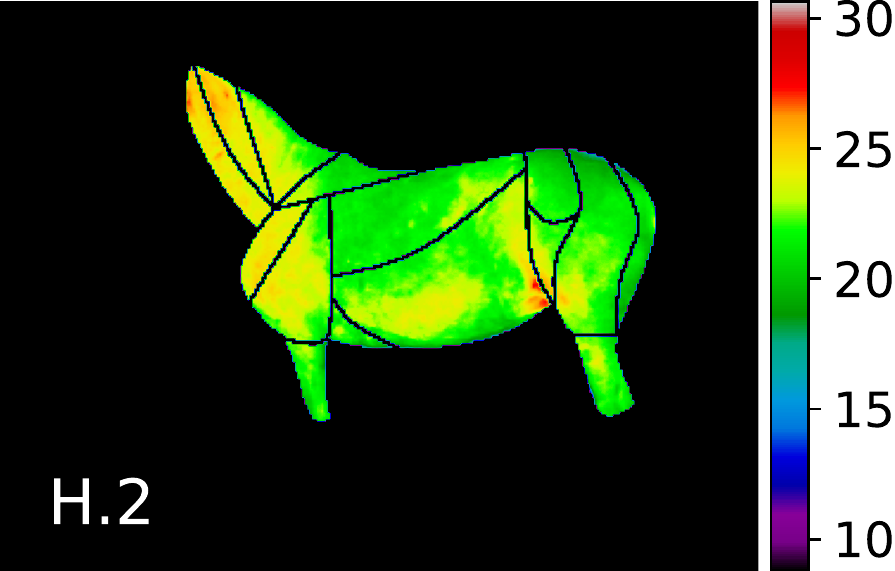}
	\includegraphics[width=0.24\linewidth]{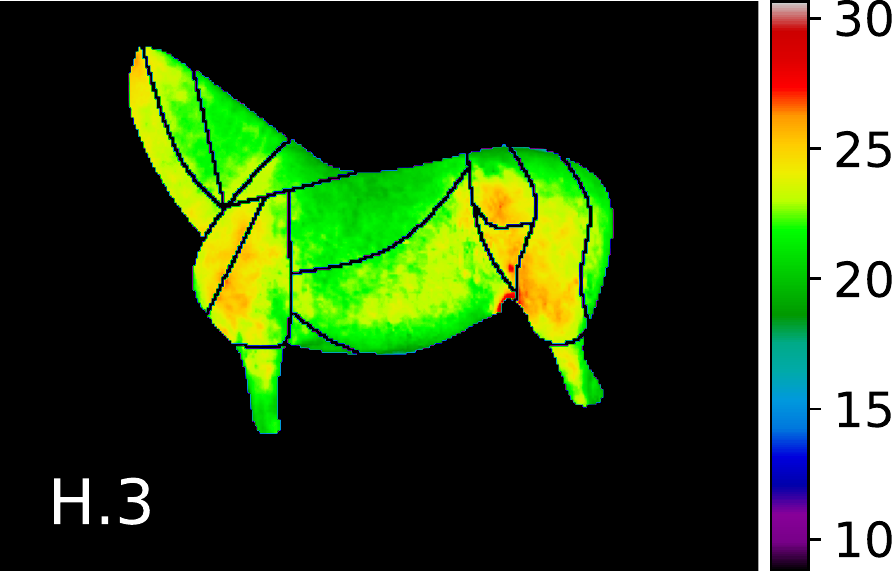}
	\includegraphics[width=0.24\linewidth]{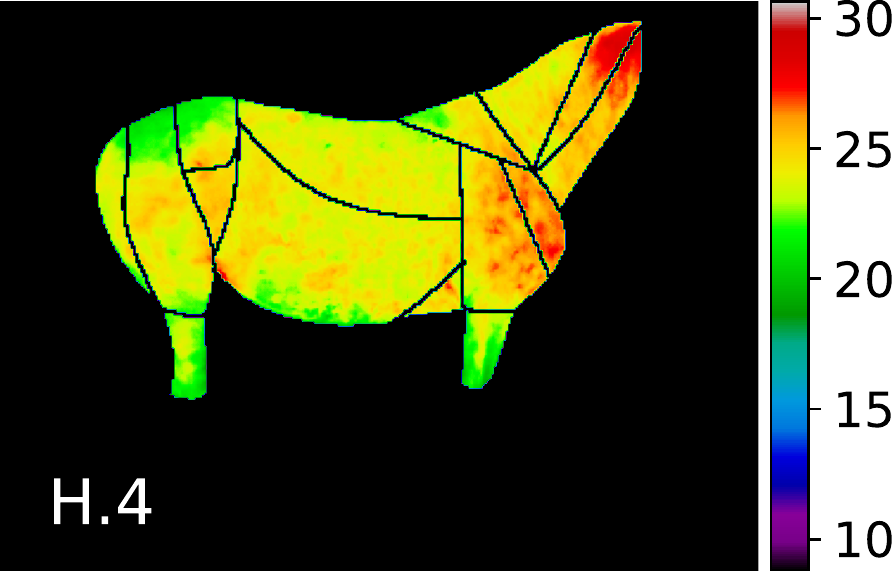}
	\includegraphics[width=0.24\linewidth]{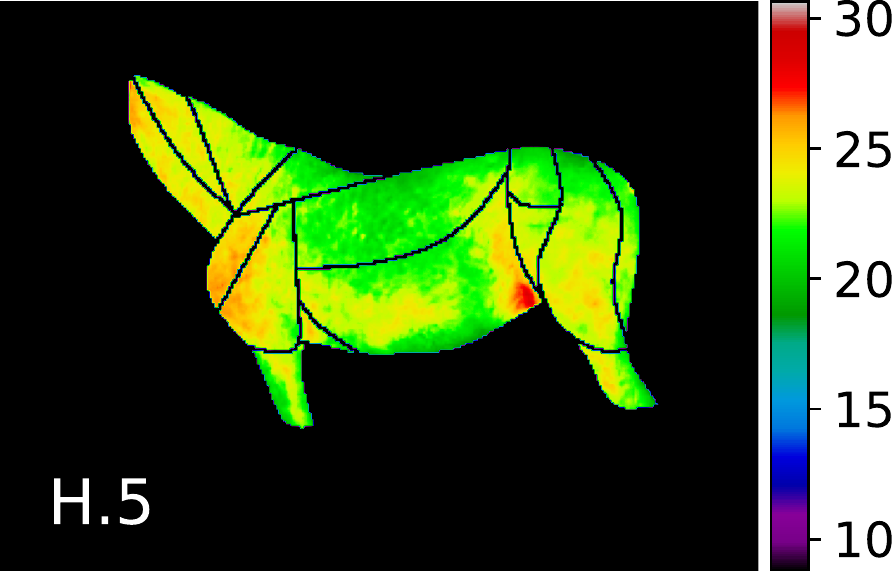}
	\includegraphics[width=0.24\linewidth]{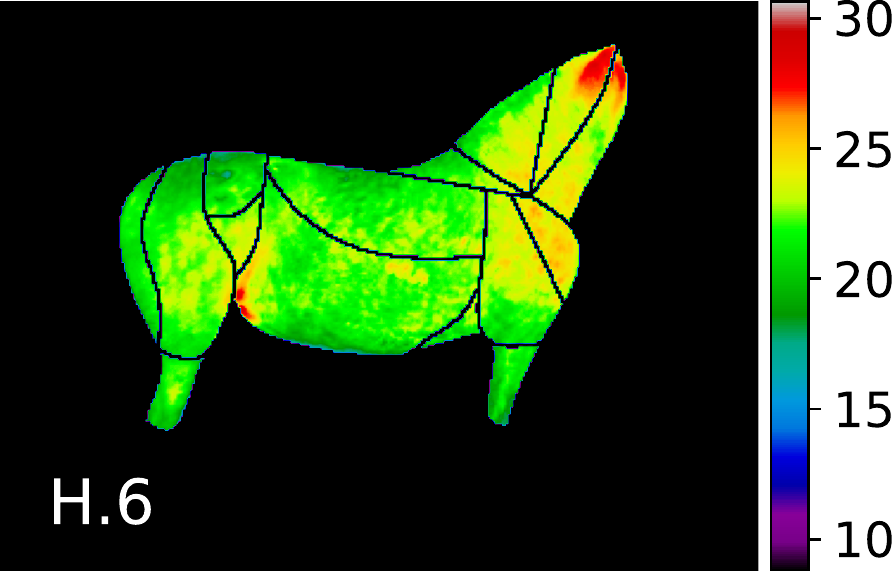}
	\includegraphics[width=0.24\linewidth]{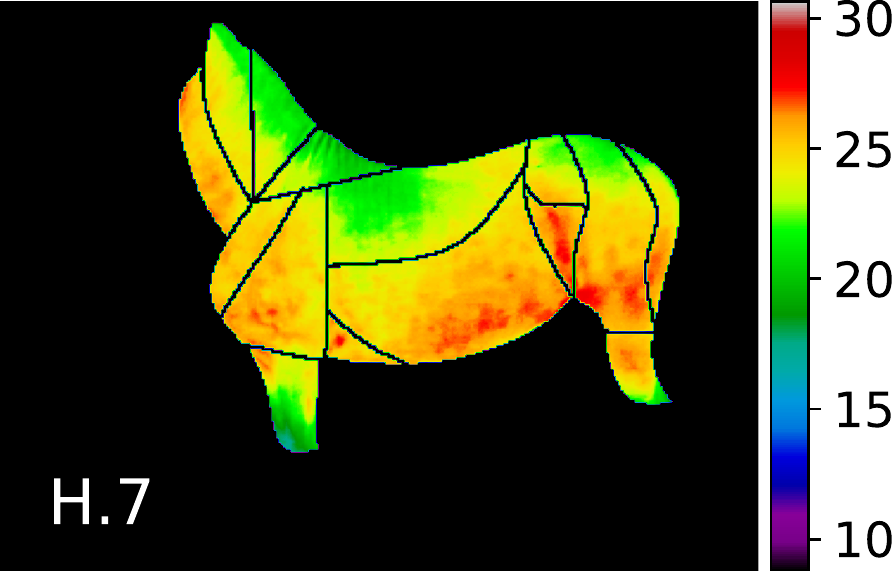}
	\includegraphics[width=0.24\linewidth]{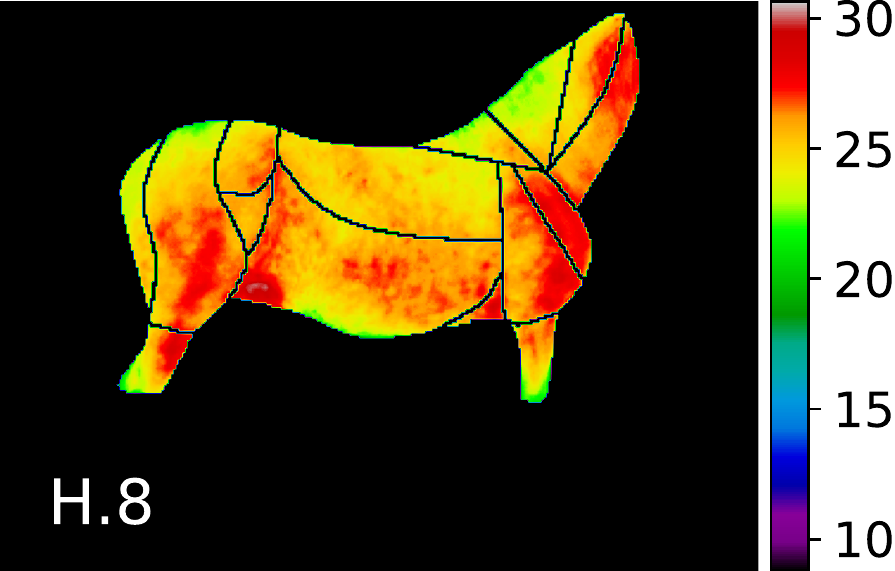}
	\includegraphics[width=0.24\linewidth]{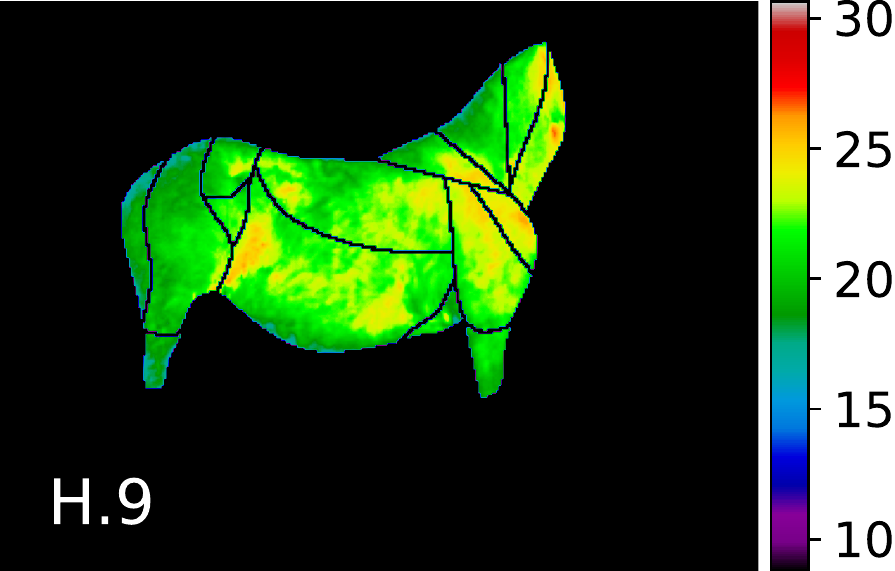}
	\includegraphics[width=0.24\linewidth]{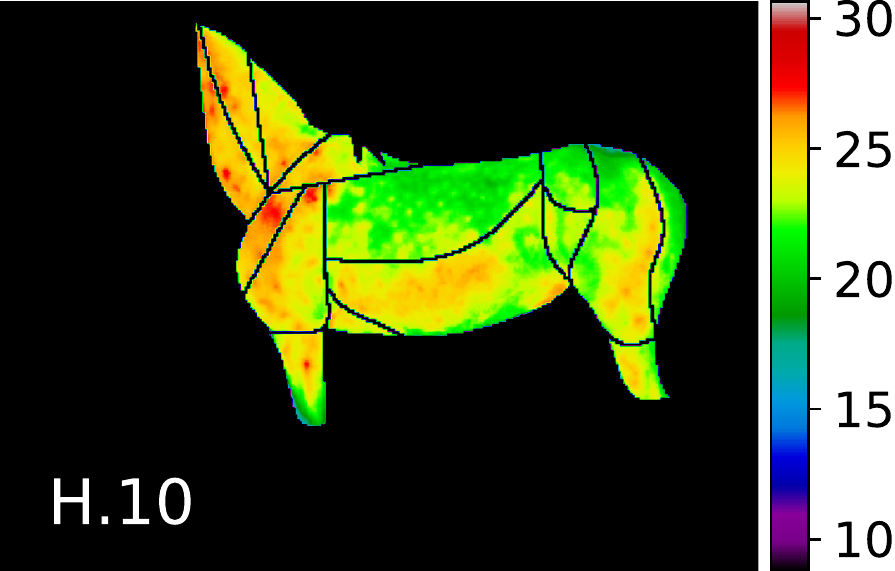}
	\includegraphics[width=0.24\linewidth]{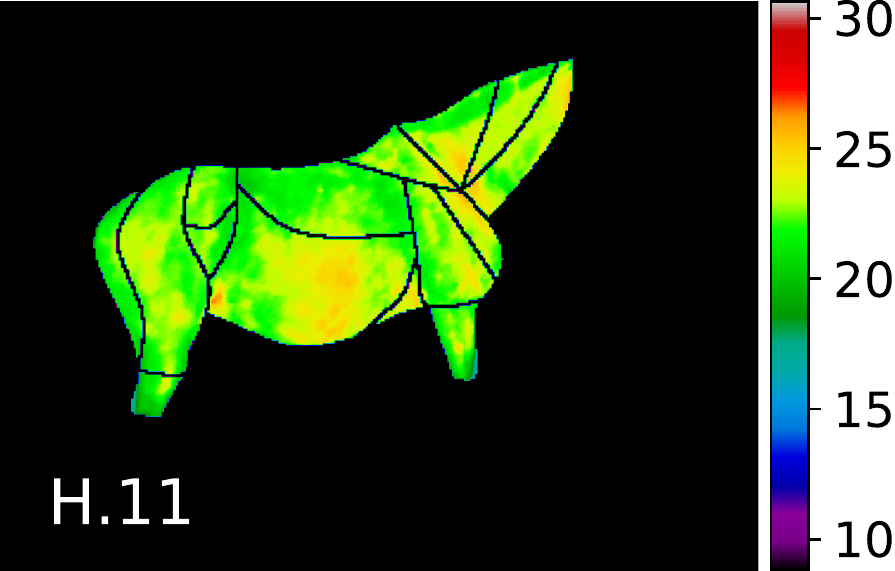}
	\includegraphics[width=0.24\linewidth]{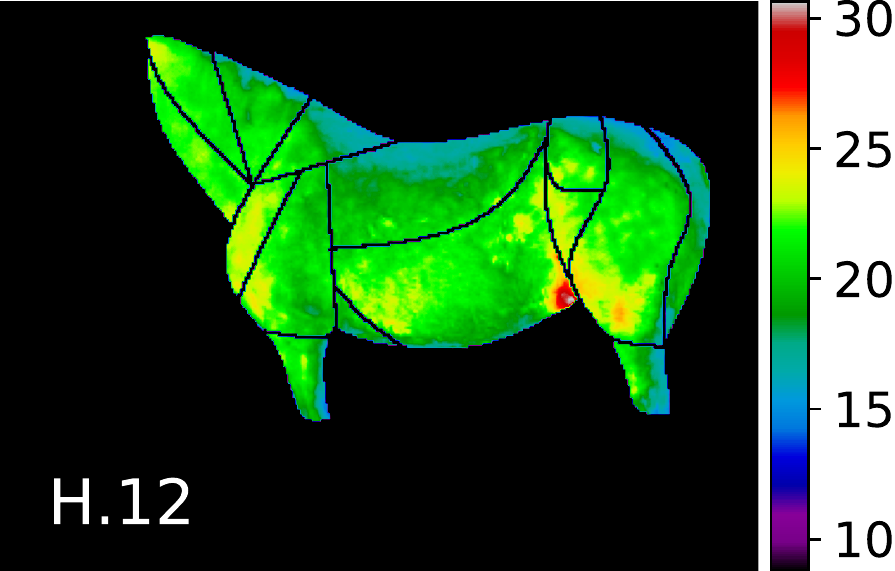}
	\includegraphics[width=0.24\linewidth]{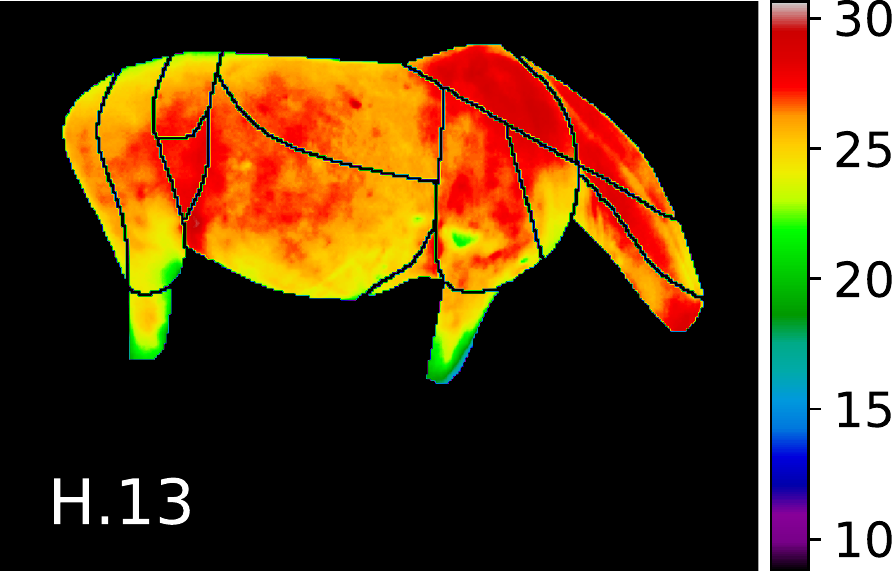}
	\includegraphics[width=0.24\linewidth]{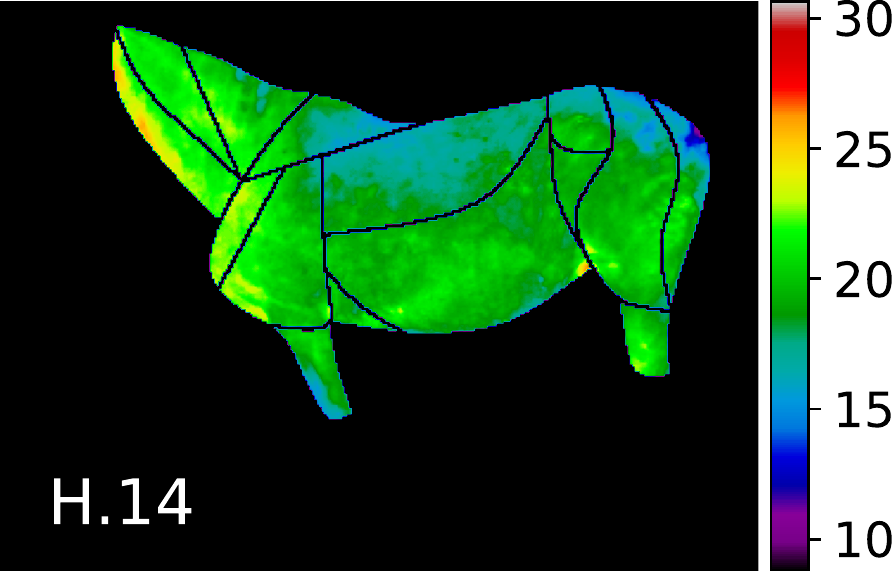}
	\includegraphics[width=0.24\linewidth]{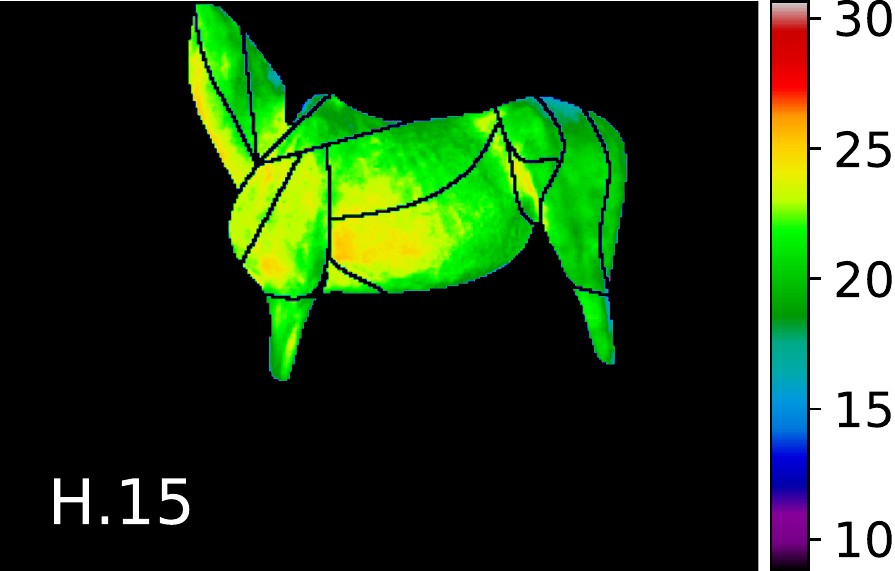}
	\includegraphics[width=0.24\linewidth]{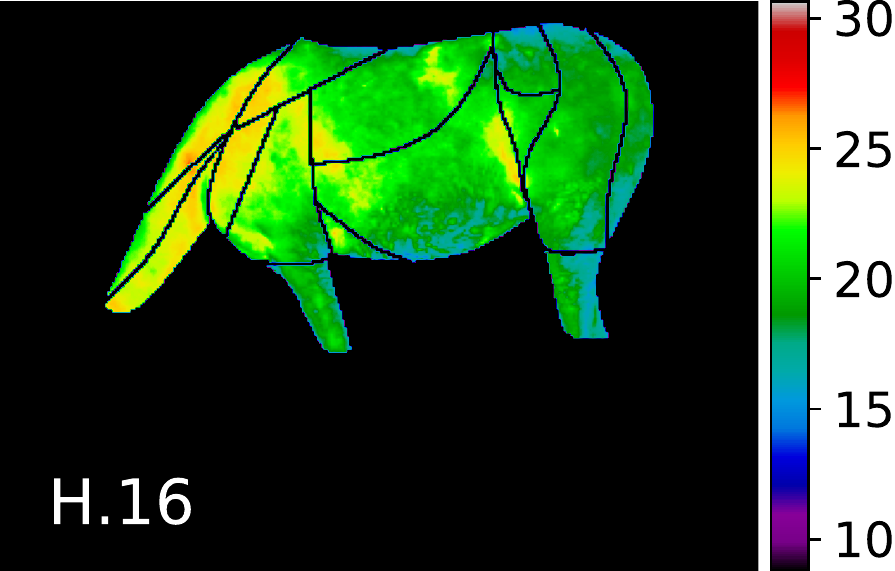}
	\caption{Thermal maps of annotated ROIs for horses in our dataset. }
	\label{fig:heatmaps_horses}
\end{figure}

\begin{figure}
	\centering
	\includegraphics[width=0.24\linewidth]{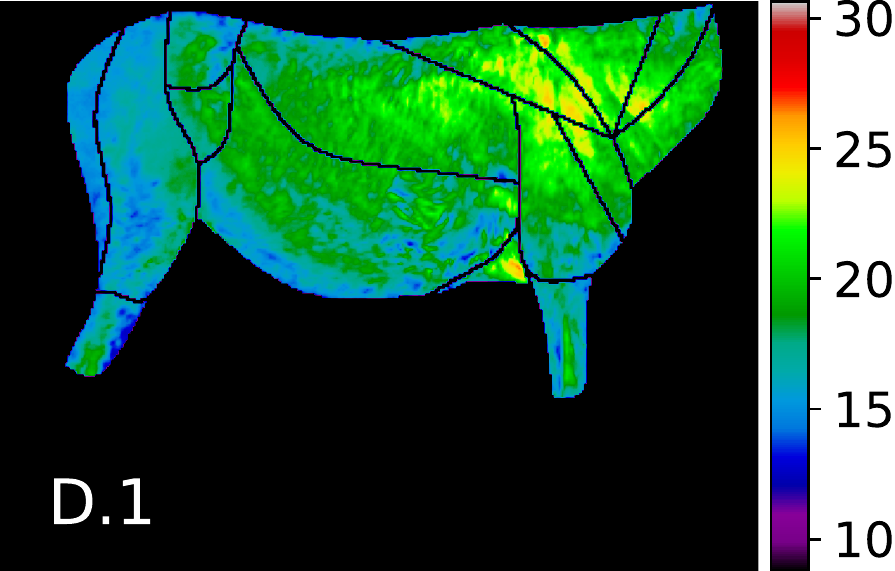}
	\includegraphics[width=0.24\linewidth]{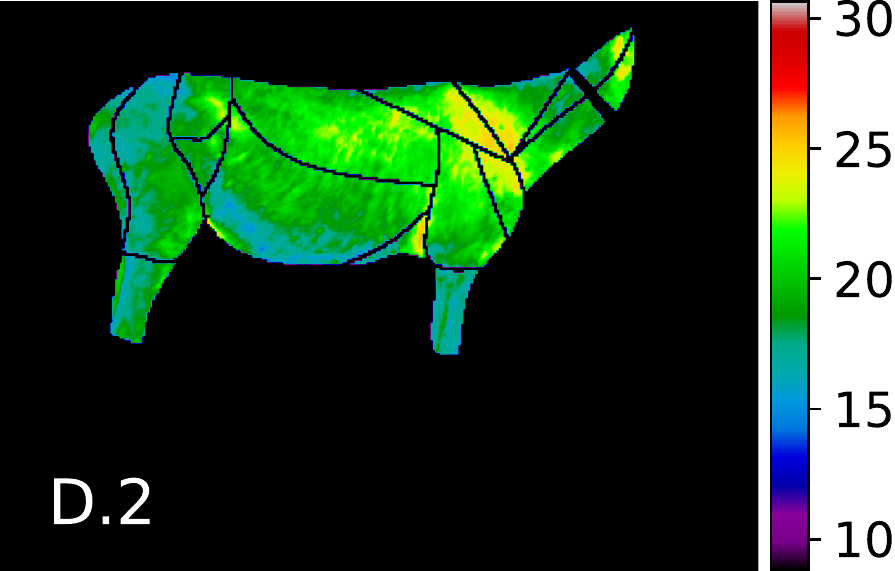}
	\includegraphics[width=0.24\linewidth]{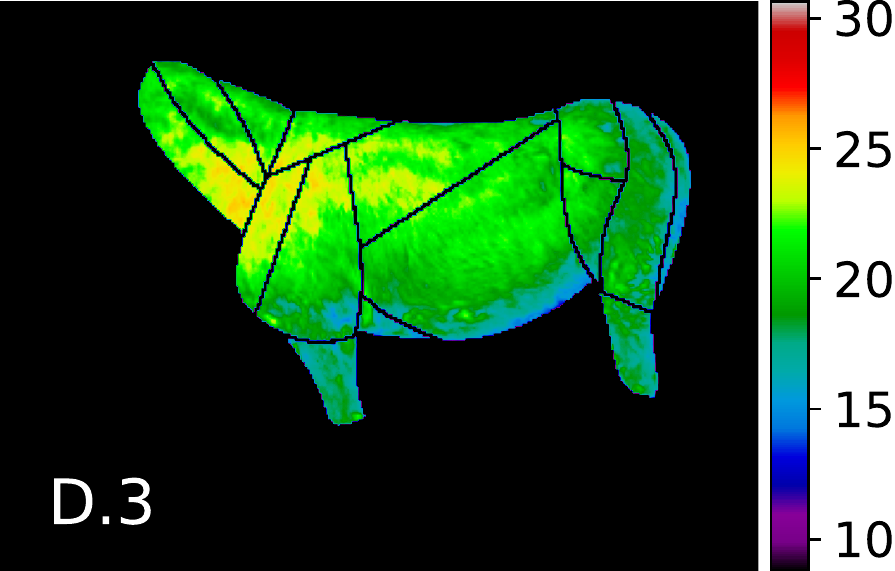}
	\includegraphics[width=0.24\linewidth]{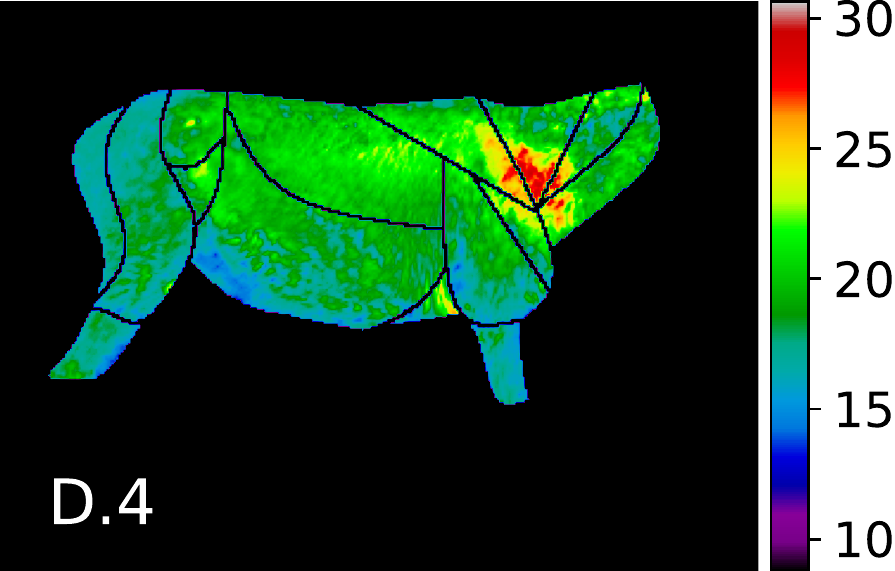}
	\includegraphics[width=0.24\linewidth]{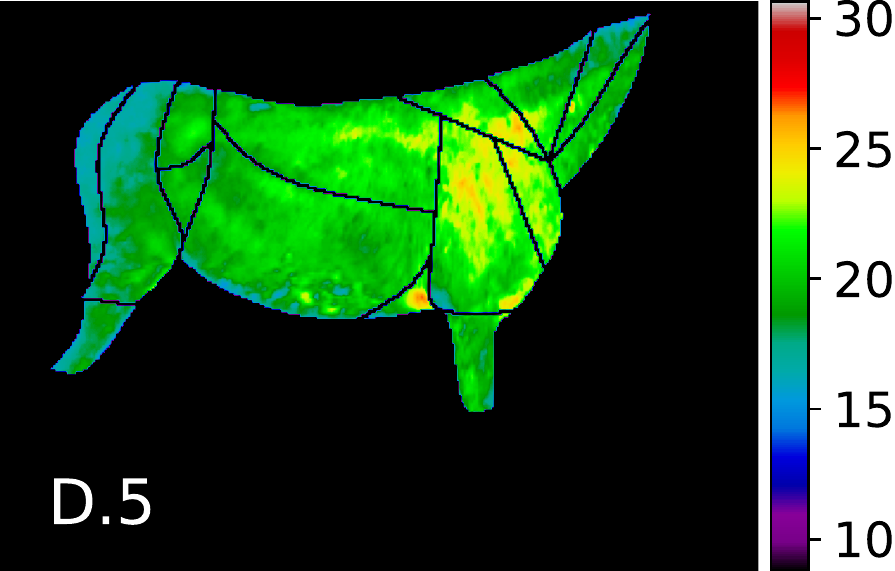}
	\includegraphics[width=0.24\linewidth]{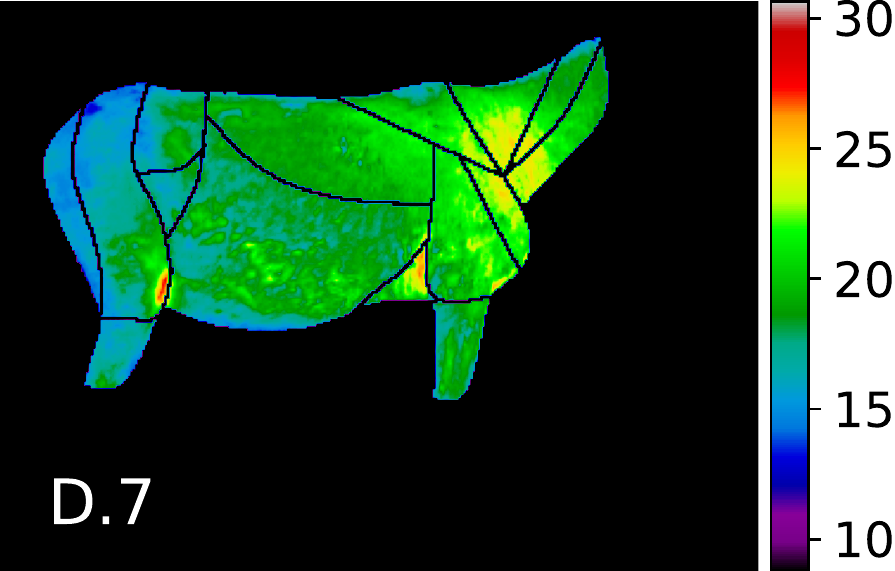}
	\includegraphics[width=0.24\linewidth]{images/heatmaps/D_7}
	\includegraphics[width=0.24\linewidth]{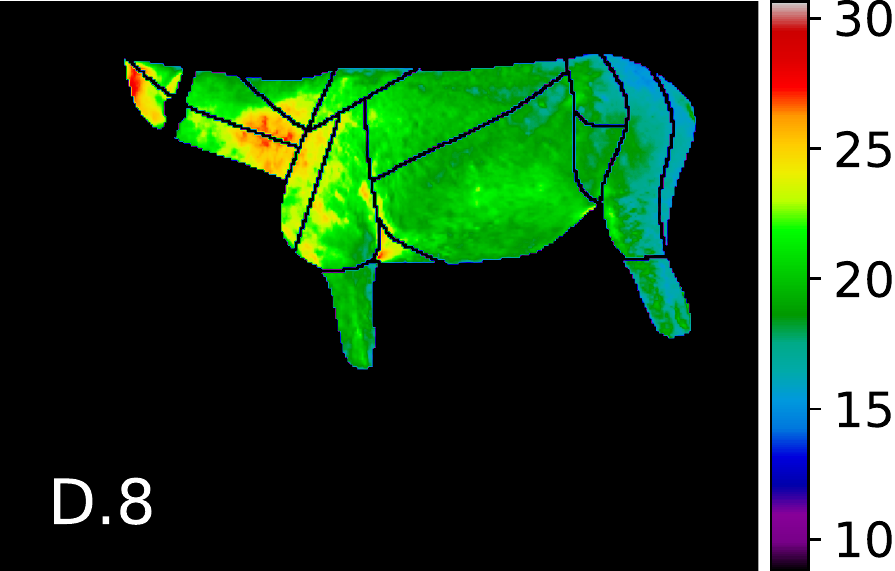}
	\includegraphics[width=0.24\linewidth]{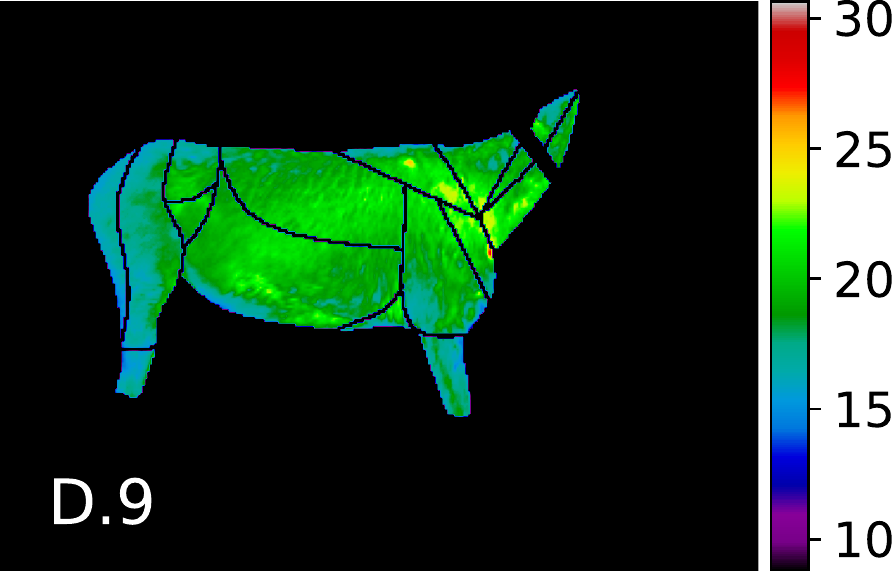}
	\includegraphics[width=0.24\linewidth]{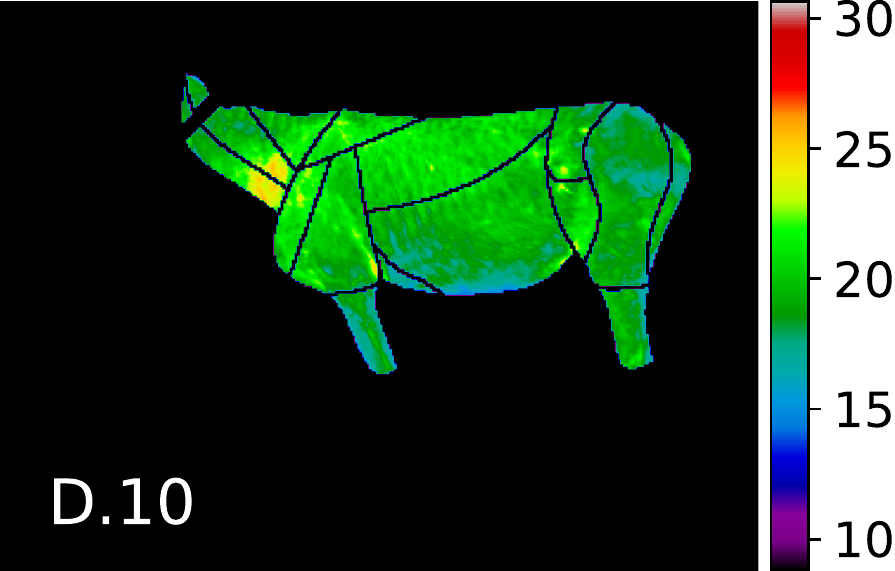}
	\includegraphics[width=0.24\linewidth]{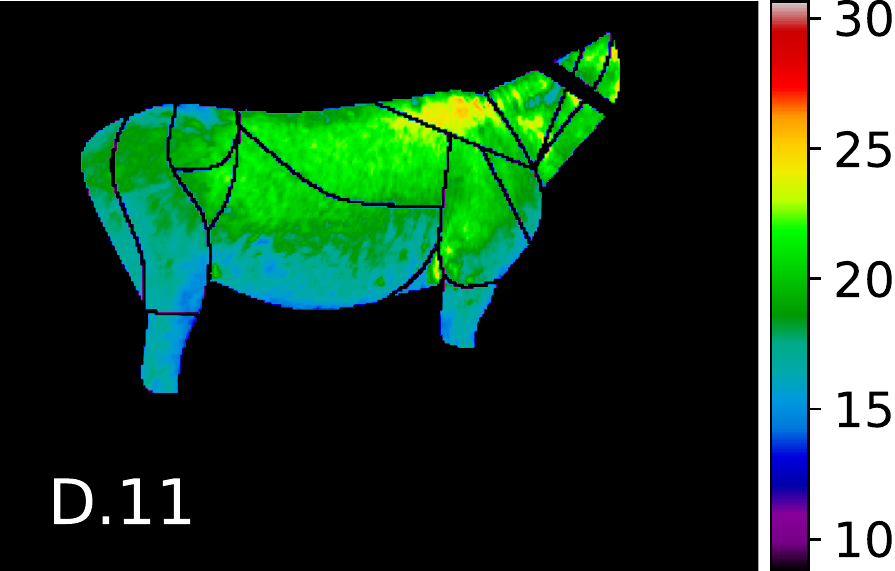}
	\includegraphics[width=0.24\linewidth]{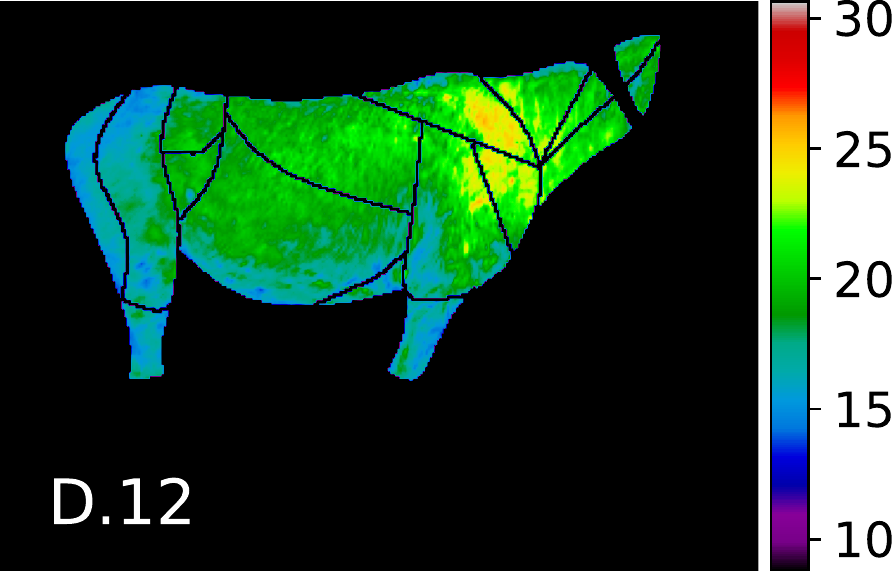}
	\includegraphics[width=0.24\linewidth]{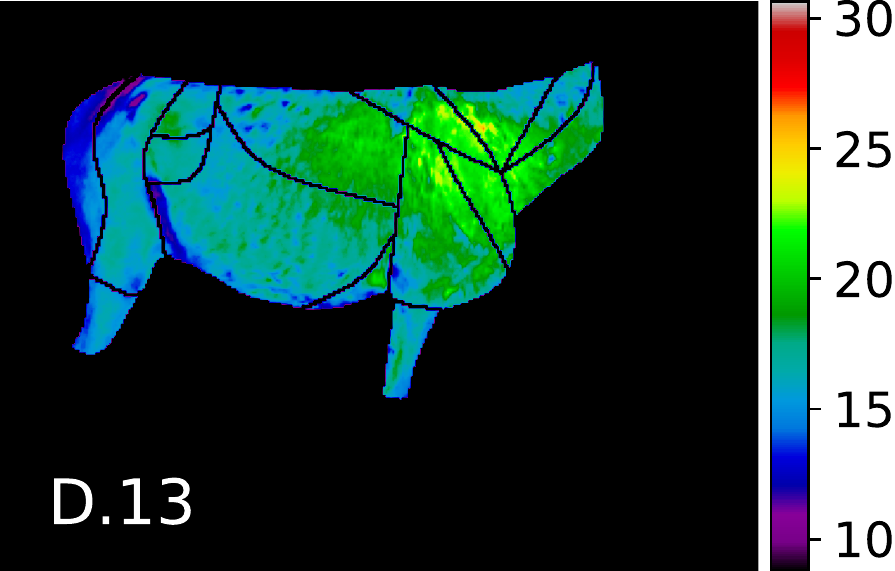}
	\includegraphics[width=0.24\linewidth]{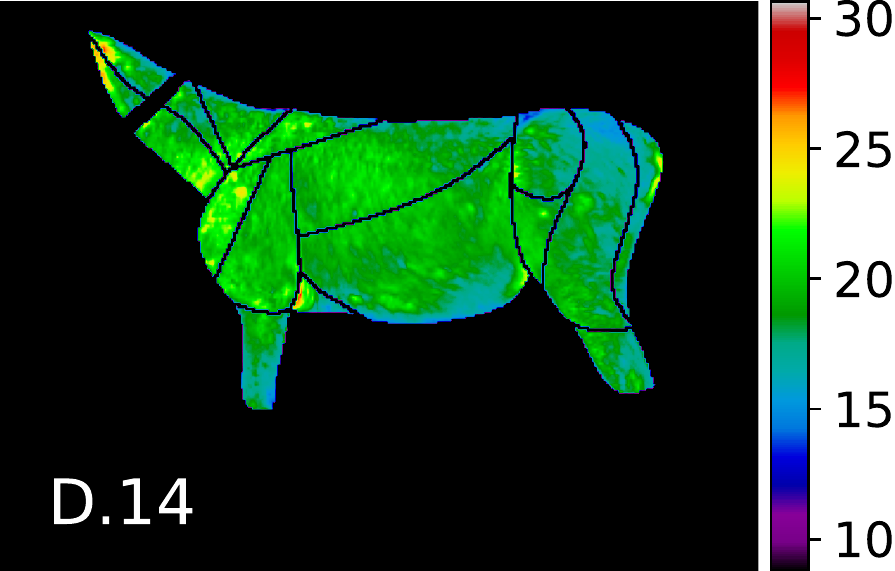}
	\includegraphics[width=0.24\linewidth]{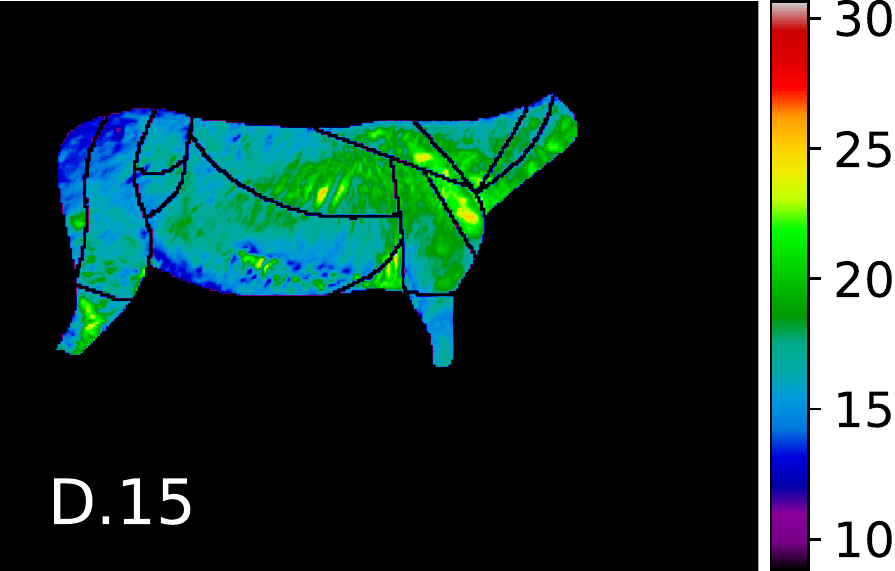}
	\includegraphics[width=0.24\linewidth]{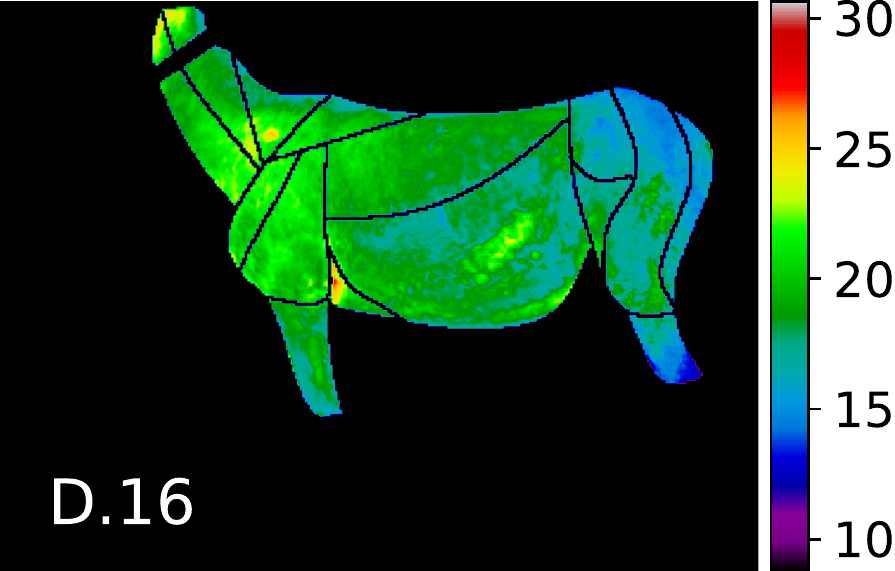}
	\caption{Thermal maps of annotated ROIs for donkeys in our dataset. }
	\label{fig:heatmaps_donkeys}
\end{figure}

Visualisation of thermal maps for horses in our study is presented in Fig.~\ref{fig:heatmaps_horses} and for donkeys in Fig.~\ref{fig:heatmaps_donkeys}. A visual comparison of the images reveals: visible temperature patterns are more complex for horses e.g. values for horses \emph{H.8, H.13} are globally higher, GOR~8 \emph{Abdomen} is warm for horses \emph{H.4, H.7, H.8, H.3} and GOR~4 \emph{Hindquarter} for horses \emph{H.4, H.7, H.8, H.10, H.13}.

\begin{figure}
	\centering
	\begin{subfigure}[b]{0.24\textwidth}
	\includegraphics[width=1.0\linewidth]{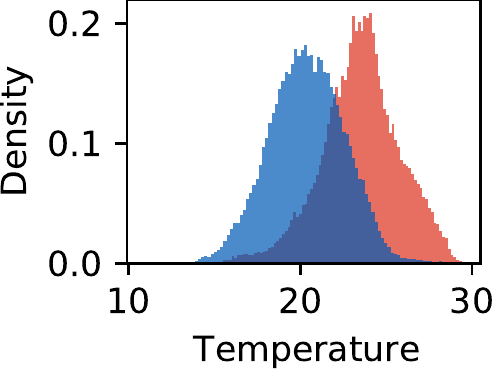}
	\caption{\emph{Frontquarter}}
	\end{subfigure}
	\begin{subfigure}[b]{0.24\textwidth}
	\includegraphics[width=1.0\linewidth]{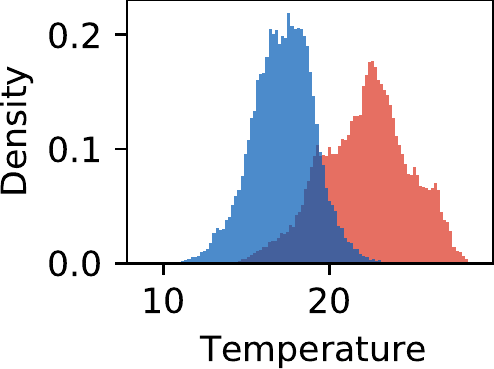}
	\caption{\emph{Hindquarter}}
	\end{subfigure}
	\begin{subfigure}[b]{0.24\textwidth}
	\includegraphics[width=1.0\linewidth]{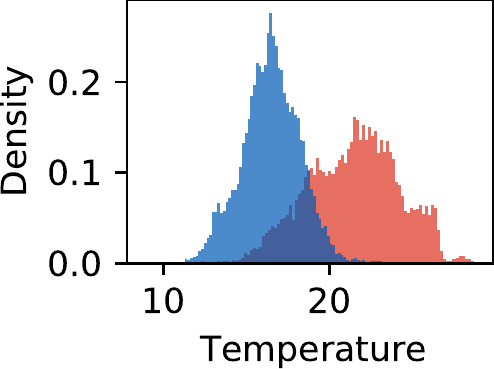}
	\caption{\emph{Rump}}
	\end{subfigure}
	\begin{subfigure}[b]{0.24\textwidth}
	\includegraphics[width=1.0\linewidth]{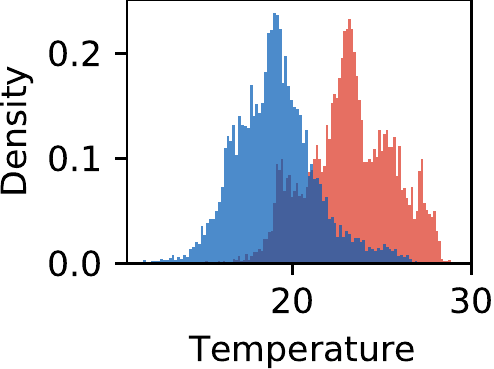}
	\caption{\emph{Groins}}
	\end{subfigure}
	\caption{Comparison of temperature histograms between animal species in  identified characteristic areas corresponding to selected groups of ROIs: (a) GOR 2 \emph{Frontquarter}; (b) GOR 4 \emph{Hindquarter}; (c) GOR 5 \emph{Rump}; (i) GOR 9 \emph{Groins}. Horses are represented in red, donkeys are blue.}
	\label{fig:group_histo}
\end{figure}

On the other hand, donkeys seem more uniform, we notice that temperature values in GOR~5 \emph{Rump} are usually lower than in other GORs, while in GOR~2 \emph{Frontquarter} we observe warm areas. A comparison of histograms for four selected GORs is presented in Fig.~\ref{fig:group_histo}. We can see that the overlap between histograms is more visible for the GOR~5 than for the GOR~2.

\begin{figure}
	\centering
	\begin{subfigure}[b]{0.49\textwidth}
	\includegraphics[width=1.0\linewidth]{images/heatmaps/H_11}
	\caption{Horse \emph{H11}, $t_c\in\langle8.8,29.6\rangle$} 
	\end{subfigure}
	\begin{subfigure}[b]{0.49\textwidth}
	\includegraphics[width=1.0\linewidth]{images/heatmaps/D_12}
	\caption{Donkey, \emph{D12}, $t_c\in\langle8.8,29.6\rangle$} 
	\end{subfigure}
	\begin{subfigure}[b]{0.49\textwidth}
	\includegraphics[width=1.0\linewidth]{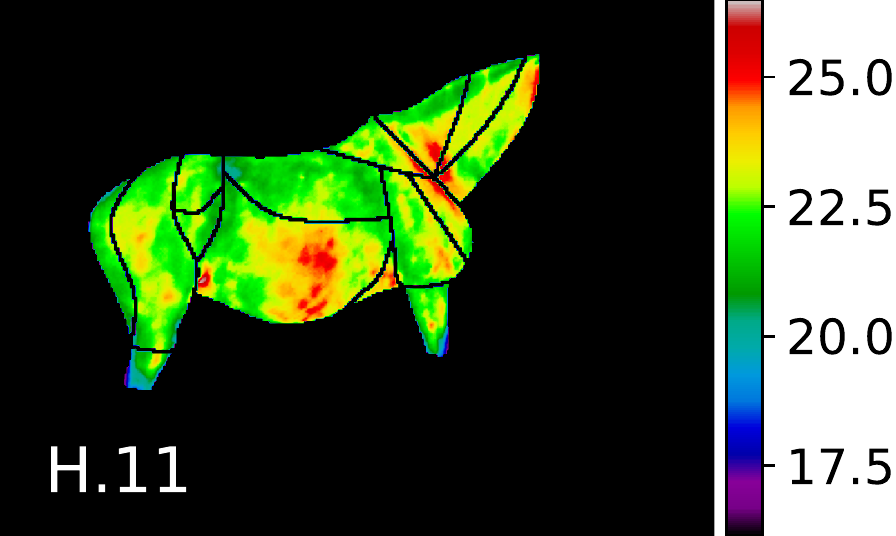}
	\caption{Horse, \emph{H11}, $t_c\in\langle13.7,26\rangle$} 
	\end{subfigure}
	\begin{subfigure}[b]{0.49\textwidth}
	\includegraphics[width=1.0\linewidth]{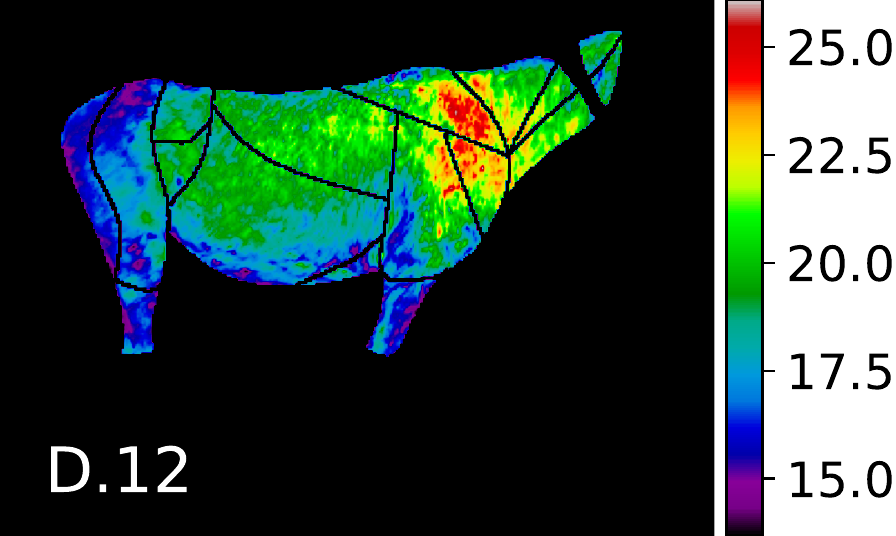}
	\caption{Donkey, \emph{D12}, $t_c\in\langle16.2,26.4\rangle$} 
	\end{subfigure}
	\caption{Selected examples of two animals from our dataset. The color map values $t_c$ for images in the upper row are scaled to the common range, which makes them easy to compare: (a) horses; (b) donkeys. Images in the bottom row are scaled to the minimal and maximal temperatures in annotated ROIs of each animal, which highlights individual thermal patterns: (c)~horses; (d)~donkeys. E.g. warm horse's GORs \emph{Abdomen} and \emph{Neck}, cool donkey's GOR \emph{Rump}, and warm donkey's GOR \emph{Frontquarter}.}
	\label{fig:heatmap_relative}
\end{figure}

To highlight the visible patterns, Fig.~\ref{fig:heatmap_relative} presents individual thermal maps for two example animals. In plots~(c) and (d) we can see that characteristic patterns are usually associated with groups of ROIs rather than an individual ROI. 

\begin{figure}
	\centering
	\begin{subfigure}[b]{0.49\textwidth}
    \includegraphics[width=1.0\linewidth]{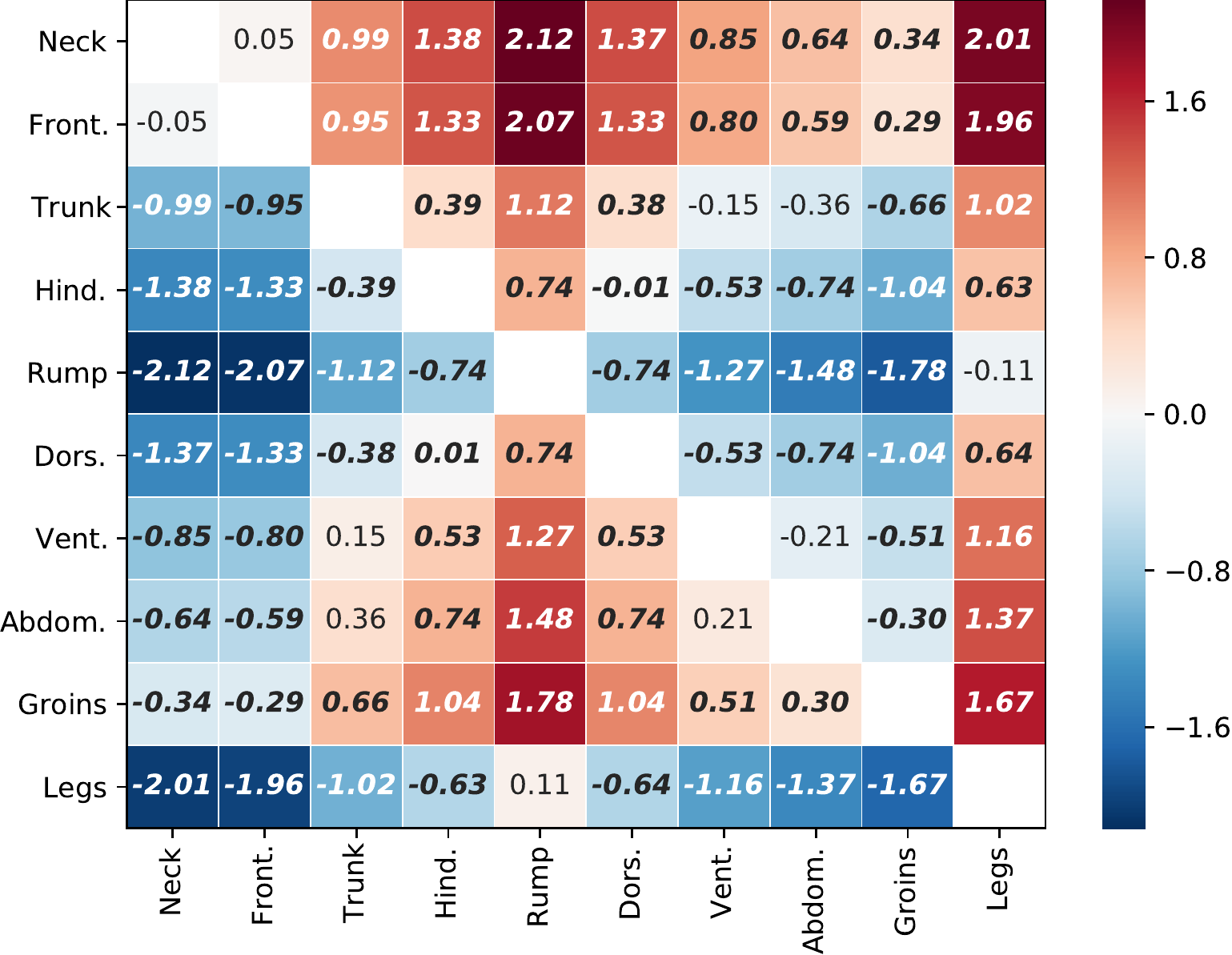}
    \caption{$\vect{M}^\Delta$,  Horses}
    \end{subfigure}
    \begin{subfigure}[b]{0.49\textwidth}
    \includegraphics[width=1.0\linewidth]{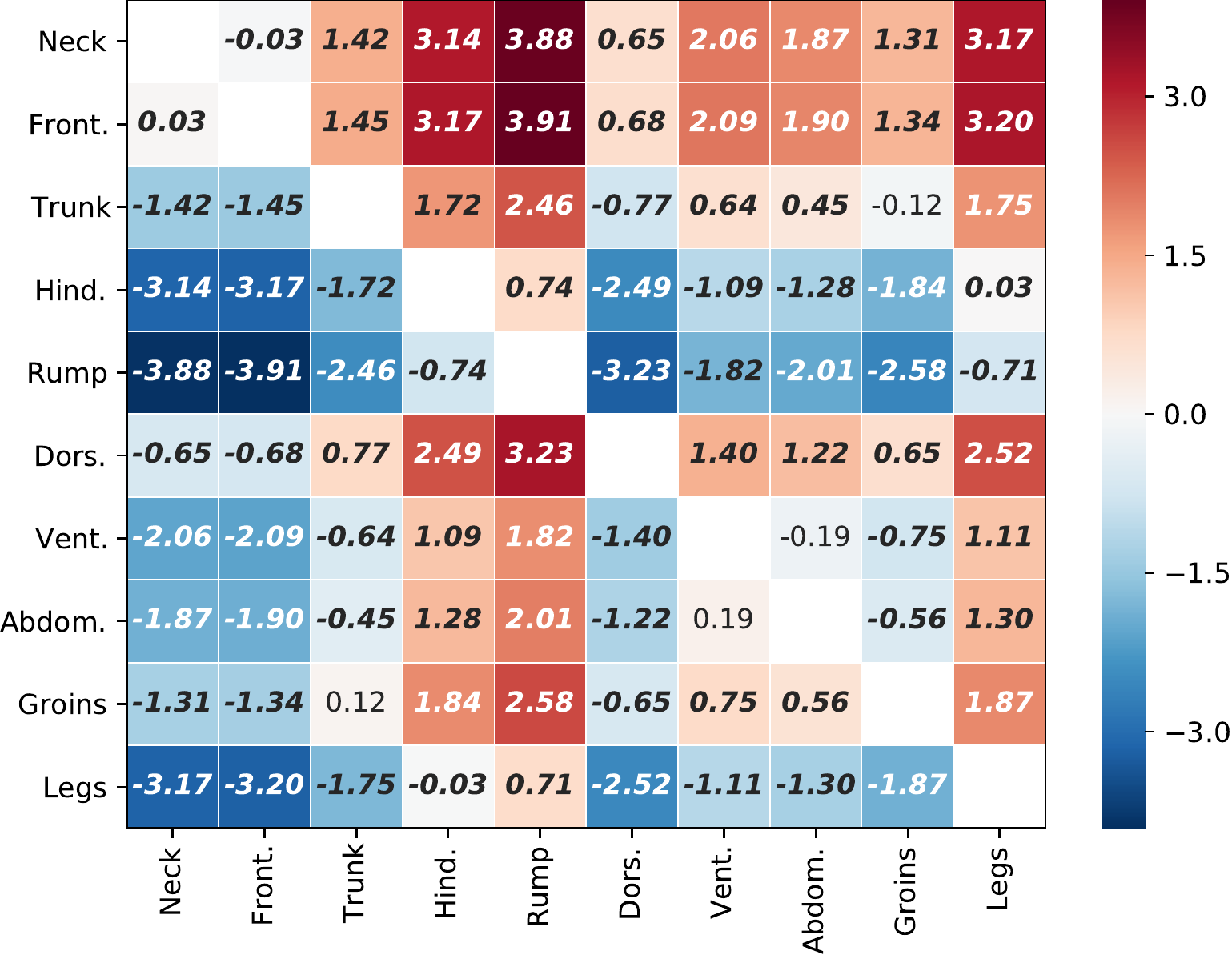}
    \caption{$\vect{M}^\Delta$, Donkeys}
    \end{subfigure}
	\begin{subfigure}[b]{0.49\textwidth}
    \includegraphics[width=1.0\linewidth]{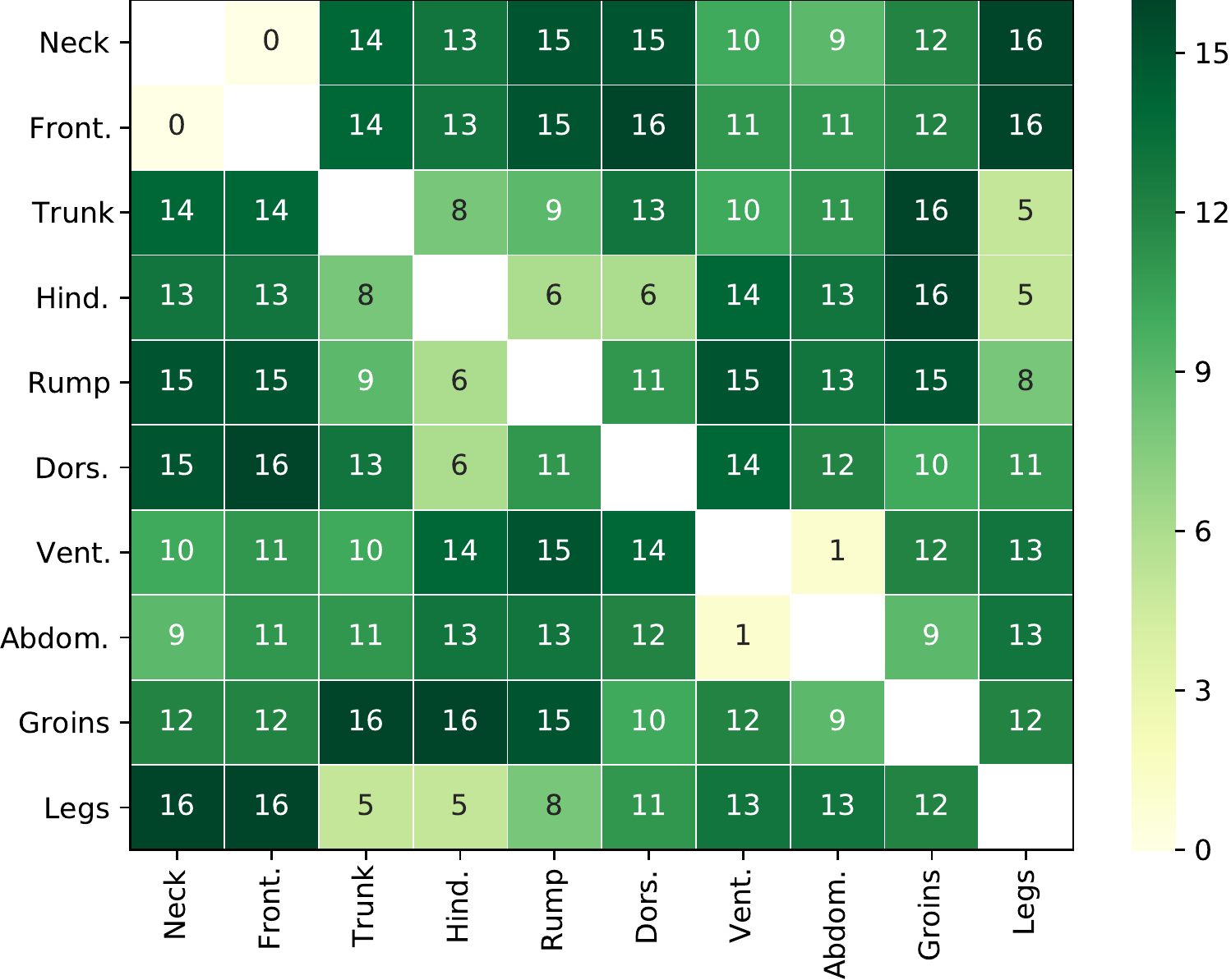}
    \caption{$\vect{M}^L$, Horses}
    \end{subfigure}
    \begin{subfigure}[b]{0.49\textwidth}
    \includegraphics[width=1.0\linewidth]{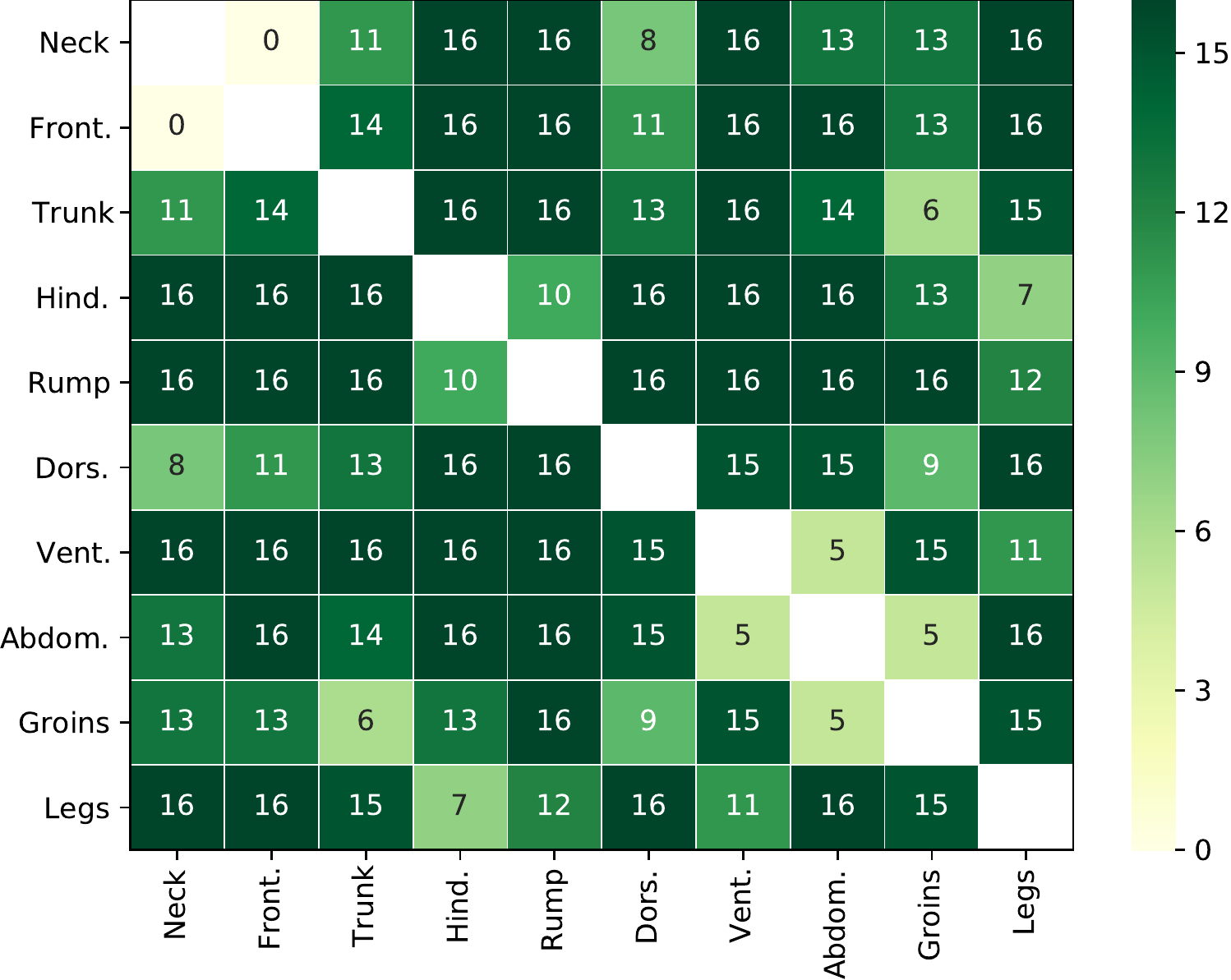}
    \caption{$\vect{M}^L$, Donkeys}
    \end{subfigure}
	\caption{Thermal patterns i.e. statistically significant differences between GORs (see Sec.~\ref{sec:thermal_patterns_method}). Upper panels present differences within one genre: (a) horses; (b) donkeys. E.g. the value $\vect{M}^\Delta_{[4,0]}=-2.12$ in the $[4,0]$ cell in the panel~(a) is the difference between mean temperatures for the pair \emph{Rump}/\emph{Neck}, indicating that the \emph{Rump} GOR is colder. Bold font indicates `global' statistical significance of this difference. Bottom panels present tables for (c)~horses (d)~donkeys, with the number of animals for which the corresponding temperature difference in the table above is statistically significant considering individual thermal pattern of this animal. E.g. the value $\vect{M}^L_{[4,0]}=15$ in panel~(c), which indicates that the pattern \emph{Rump}/\emph{Neck} is locally significant for $15$ horses. Note that the most stable patterns should be statistically significant simultaneously for all data combined and for each of the 16 animals of a given species.
}	\label{fig:matrices}
\end{figure}

Thermal patterns for both species i.e. differences of temperatures between designated GORs, prepared using the methodology described in Sec.~\ref{sec:thermal_patterns_method} are presented in Fig.~\ref{fig:matrices}. For both species, GORs \emph{Rump} and \emph{Legs} are consequently colder that others while GORs \emph{Neck} and \emph{Frontquarter} are warmer. The majority of differences are globally significant, for horses there are five exceptions: \emph{Neck}/\emph{Frontquarter},  \emph{Trunk}/\emph{Ventral aspect}, \emph{Trunk}/\emph{Abdomen}, \emph{Ventral aspect}/\emph{Abdomen} and \emph{Rump}/\emph{Legs}. For donkeys there are only two exceptions: \emph{Ventral aspect}/\emph{Abdomen} and \emph{Trunk}/\emph{Groins}.

However, as for the local significance of differences: for horses there are only five patterns that consequently appear for all animals: 
\emph{Dorsal aspect}/\emph{Frontquarter}, \emph{Groins}/\emph{Trunk}, \emph{Groins}/\emph{Hindquarter}, \emph{Legs}/\emph{Neck}, \emph{Legs}/\emph{Frontquarter}. On the contrary, for donkeys, there are 21 of such patterns, which indicates that donkeys are individually more consistent with the global trend, which will be further addressed in the discussion. 

\begin{figure}
	\centering
	\begin{subfigure}[b]{0.49\textwidth}
    \includegraphics[width=1.0\linewidth]{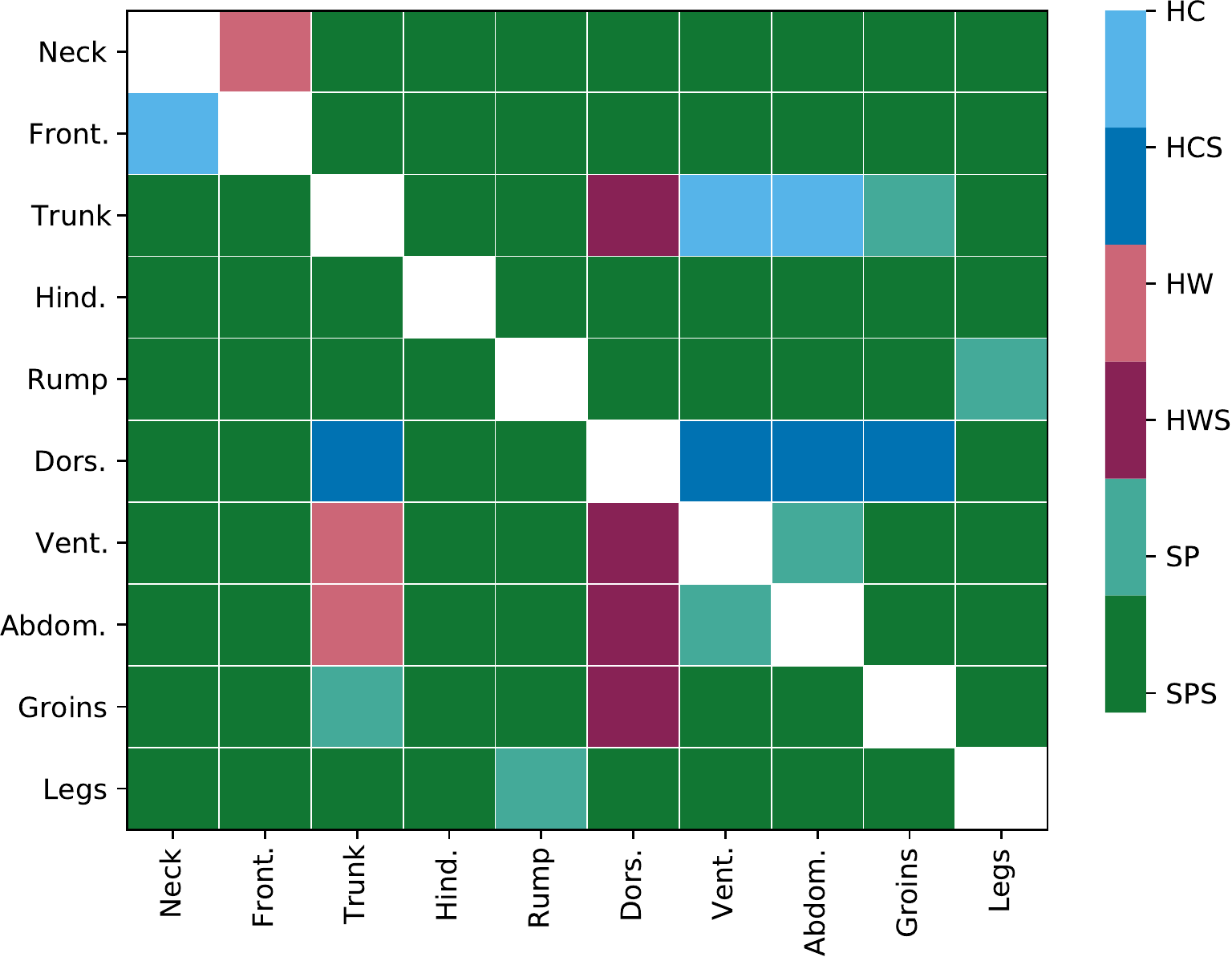}
    \caption{Global similarity of patterns}
    \end{subfigure}
    \begin{subfigure}[b]{0.49\textwidth}
    \includegraphics[width=1.0\linewidth]{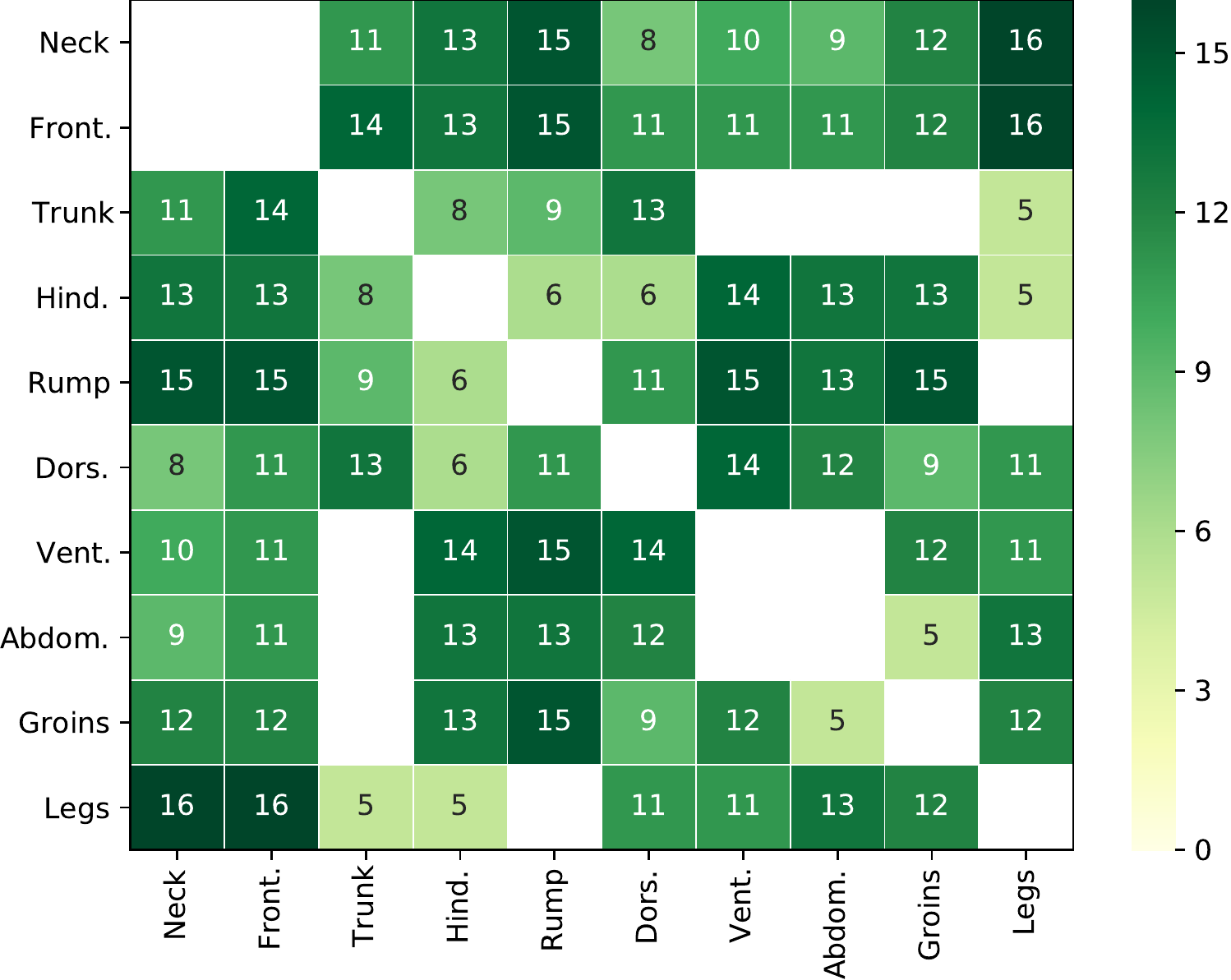}
    \caption{Min. no. animals with a given pattern}
    \end{subfigure}
	\caption{Comparison of thermal patterns for both species: (a) division of thermal patterns into six classes: SPS -- denotes thermal patterns that are similar and globally statistically significant for both species; SP: similar patterns but not significant; HWS: opposite patterns where horses are warmer (and donkeys colder), which are statistically significant; HW: same as HWS but not significant; HCS: significant patterns where horses are colder (and donkeys warmer); HC: same as HCS but not significant. Note that the HWS class is the most common which suggests global similarity of patterns; (b) the minimum number of animals that confirm the global trend for classes $\{SPS, HWS, HCS\}$ i.e. for both species at least this number of animals share a given pattern. Note that the patterns where this value is high (e.g. 16, which is the maximum) are more reliable.}
	\label{fig:matrices_comp}
\end{figure}

A summary of pattern similarities between both species is presented in Fig.~\ref{fig:matrices_comp}. Panel~(a) presents patterns which are similar for both species e.g. the relation in temperatures between GORs \emph{Rump}/\emph{Neck} is the same for both species (the GOR \emph{Rump} is colder than the GOR \emph{Neck}) and is globally, statistically significant, which is indicated with the green colour (the SPS class) in the image. Panel~(b) presents the minimal number of animals in each species that share the corresponding pattern, e.g. for the pair \emph{Rump}/\emph{Neck},  there are $15$ horses and $15$ donkeys for which the pattern is also locally, statistically significant. We can see that the majority of patterns fall under the SPS class which supports the thesis about similarities in patterns for both species. In addition, we notice that dissimilar patterns are most common in GORs \emph{Dorsal aspect} and \emph{Trunk}. The possible explanation of this observation will be discussed in the next section.

\section{Discussion}
Surface temperatures in horses are, on average, higher than in donkeys and their individual temperatures vary more within the species. This is largely due to the differences in the thermal properties of the skin and hair coat. Both, a subcutaneous fat plus skin thickness and the length of hair coat were higher in donkeys than in horses (see values in Tab~\ref{tab:animalsft}), which may indicate that they provide better thermal insulation. Recent results suggests that the hair coat properties of donkeys and horses differed significantly \cite{osthaus2018hair}. Although the authors indicate, contrary to us, a lower hair length in donkeys than in horses. This may be due to the considerable large seasonal variation in hair weight and length typical for horses, but not for donkeys, or different breeds of horses participating in our research (warmblood horses/ponies) and theirs (UK-native cold blood horses/ponies)\cite{osthaus2018hair}. In other recent studies, the strong relationship between BCS (body condition score) and SF-Skin, for both donkeys and horses, were demonstrated \cite{quaresma2013relationship, silva2016relationships}. In our research, the higher SF-Skin thickness in donkeys than in horses may indicate greater adiposity of donkeys and thus better isolation. As a result, slight local changes in donkey surface body temperature may be difficult to observe. This makes the warm area visible in regio scapularis associated with the GOR~2 \emph{Frontquarter} particularly interesting. Additionally, it suggests the validity of animal temperatures analysis through comparing the characteristics of different regions of a given animal.

\subsection{Similarities in thermal patterns of horses and donkeys}
We determined that patterns in IRT images from our data set are often visible in groups of ROIs and proposed a methodology of assessing these patterns based on the difference of temperatures in groups. Visualisation in Fig.~\ref{fig:matrices_comp} shows that these differences are similar for both species: looking at the plot~(a), we can see that $77.8\%$ of patterns are similar and statistically significant, $8.9\%$ of patterns are opposite and the rest of them cannot be confirmed statistically based on our data. We also see that for $88.8\%$ of globally significant patterns, half or more individual animals from every species share this pattern.
In our opinion, this supports the thesis about similarities in IRT images of both species. 

Analysing the values in Fig.~\ref{fig:matrices} we can see that the observation that donkeys are more `uniform' in their GORs, results in larger maximum differences between GORs and the fact that more individual animals share the global trend then for horses.

As for opposite thermal patterns, we notice that they are usually associates with GORs: \emph{Dorsal aspect} and \emph{Trunk}. Both groups cover a relatively large area of the animal's body, which raises the question of whether the more granular segmentation of these areas will show further similarities.

An important question is, whether the trends observed for our data set are characteristic of entire populations. As the number of cases is limited by practical considerations, we believe that our results should be treated as a significant indication of the existence of the relationships we described. At the same time, we emphasize the need to further verify these conclusions for more data. To facilitate this, all research data related to this study is available to the public under open licenses.

\subsection{Special cases}
\begin{figure}
	\centering
	\begin{subfigure}[b]{0.49\textwidth}
    \includegraphics[width=1.0\linewidth]{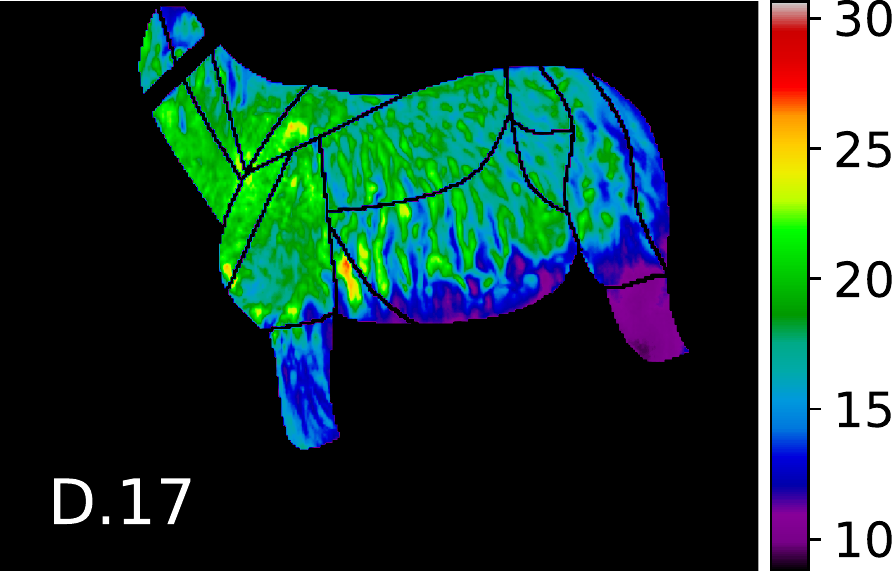}
    \caption{\emph{D.17}}
    \end{subfigure}
    \begin{subfigure}[b]{0.49\textwidth}
    \includegraphics[width=1.0\linewidth]{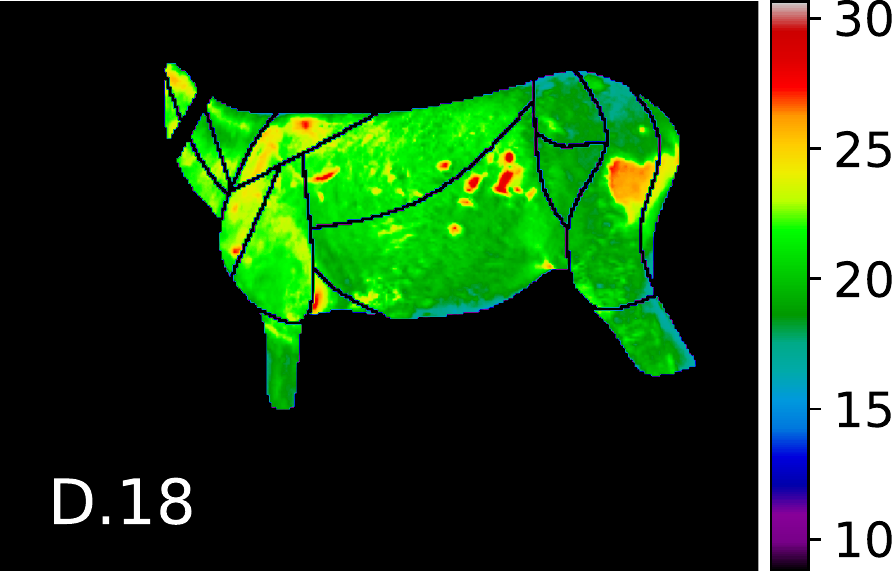}
    \caption{\emph{D.18}}
    \end{subfigure}
	\begin{subfigure}[b]{0.49\textwidth}
    \includegraphics[width=1.0\linewidth]{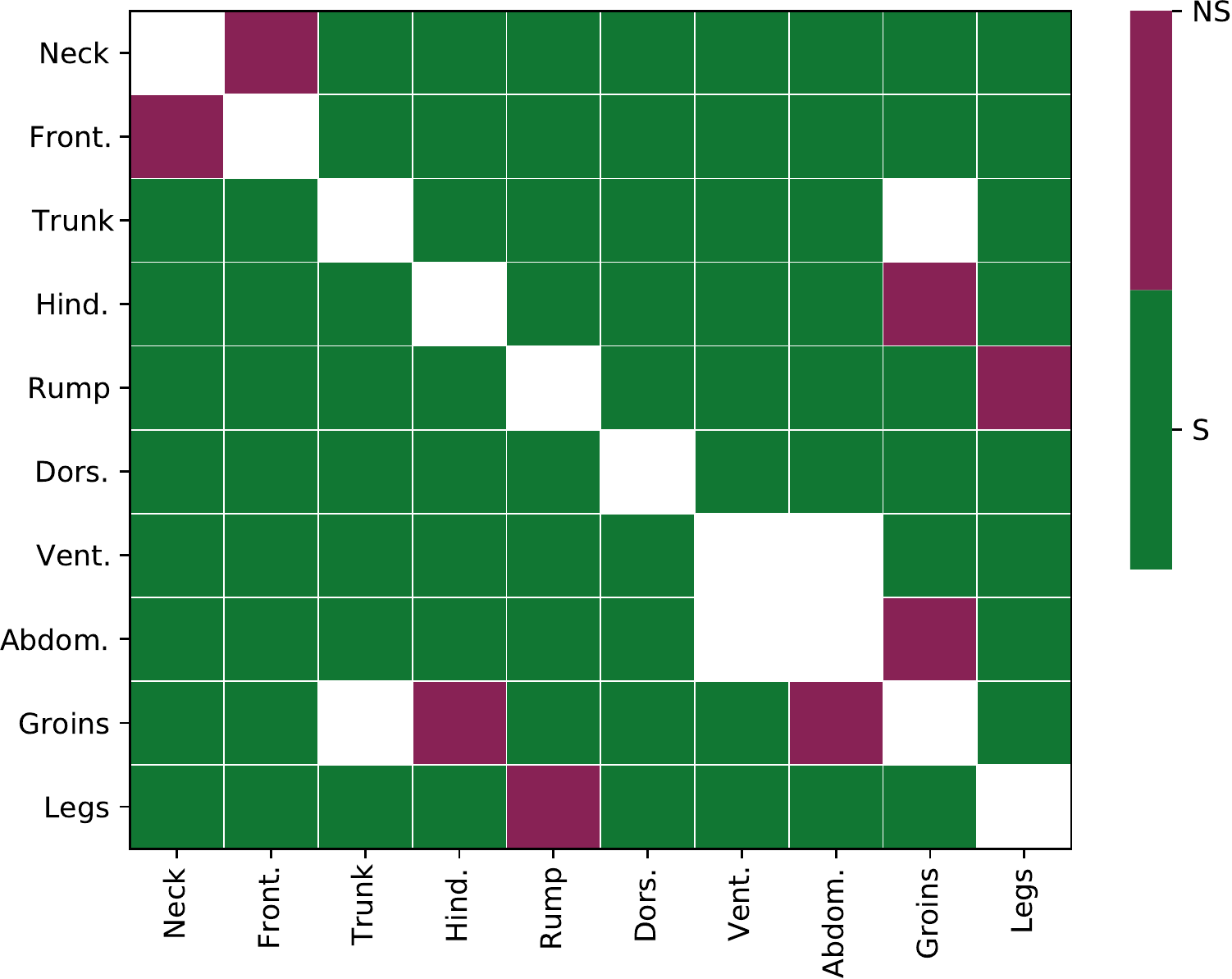}
    \caption{\emph{D.17} and Donkeys}
    \end{subfigure}
    \begin{subfigure}[b]{0.49\textwidth}
    \includegraphics[width=1.0\linewidth]{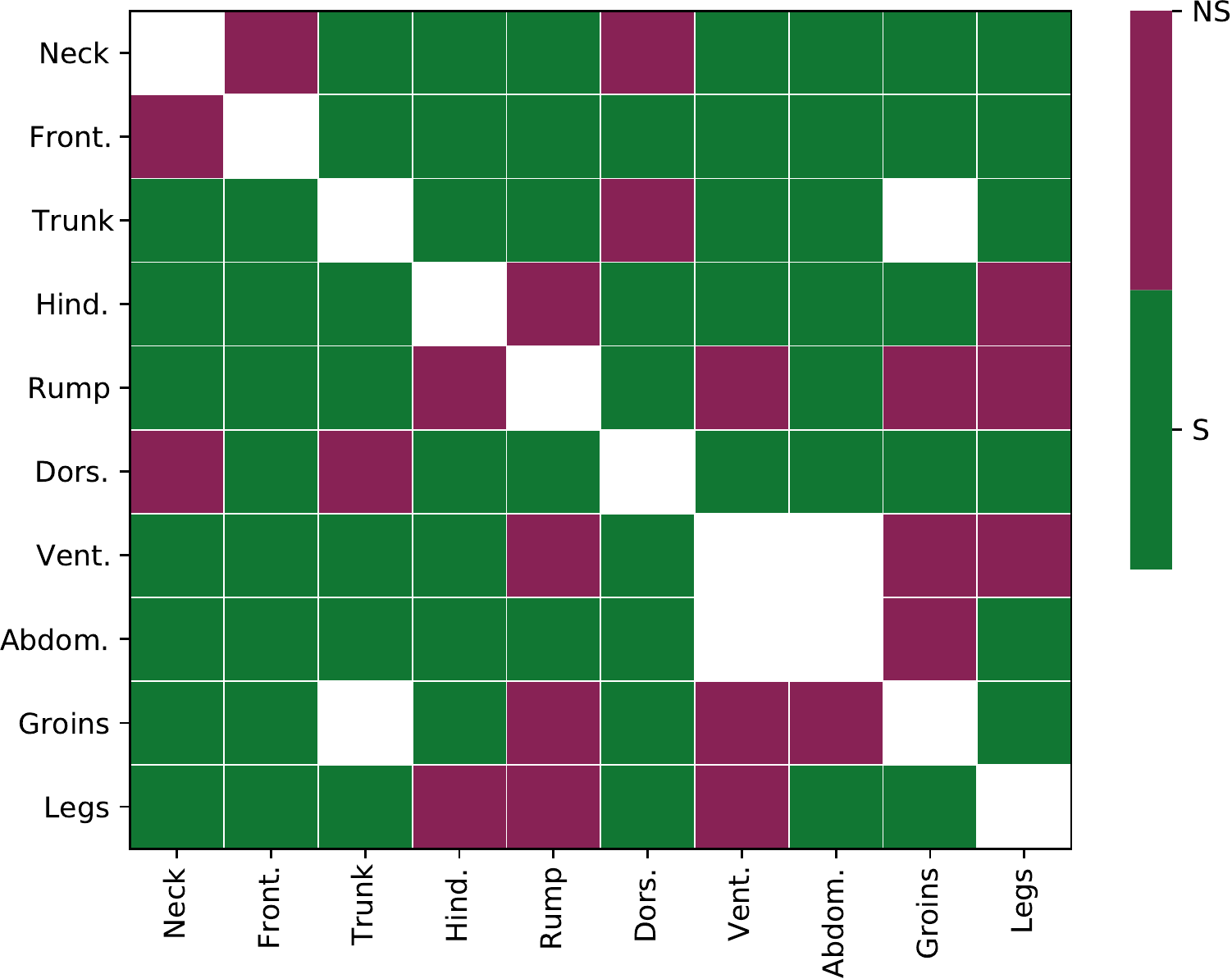}
    \caption{\emph{D.18} and Donkeys}
    \end{subfigure}
	\caption{Visualisation of differences between donkeys \emph{D.17} and \emph{D.18}, which were identified as outlier cases (see Sec.~\ref{sec:animals}), and the rest of the donkeys i.e. animals \emph{D.1-16}. The upper panels present thermal maps of the two cases: (a) Donkey \emph{D.17}; (b) Donkey \emph{D.18}. Notice that \emph{D.17} is colder than other animals due to its long hair length and that \emph{D.18} has an unusual pattern of warm areas resulting from plaque loses in the hair coat. Bottom plots show differences in their thermal patterns compared to the global pattern of other donkeys: (c) Donkey \emph{D.17} compared to Donkeys; (d) Donkey \emph{D.18} compared to Donkeys. The \emph{S} class (green) indicates that the individual animal pattern is in line with the global trend, class \emph{NS} (red) indicates the opposite.}
	\label{fig:matrices_outliers_comp}
\end{figure}

The two animals identified as outliers, i.e. \emph{D.17-18} allow for an interesting study of how general or specific our proposed thermal patterns are. The visualization of the differences between the patterns of these animals and the rest of the donkeys is presented in Fig.~\ref{fig:matrices_outliers_comp}. In panels~(b, c) the green color indicates the compliance of the animal thermal pattern with the global trend, i.e. the difference for a given pair of GORs has the same sign and is statistically significant for the animal. As we can see, thermal patterns for the \emph{D.17} donkey with a thick hair coat are usually in line with the global pattern, except for four pairs of GORs, where the statistical significance of the local animal pattern could not be confirmed. 
The patterns of the donkey \emph{D.18} are much less in line with the global pattern. This is an expected result, as the losses in the hair coat visibly affect its thermal characteristics in the image. This also suggests that the proposed thermal patterns may be the basis for creating temperature indexes as e.g. in~\cite{soroko2019evaluation} or features for detecting anomalies. Therefore, we speculate that a lot of research
methods applied successfully in equine veterinary medicine can also be used for donkeys IRT imaging, provided that the visualization conditions described by us are met.

The temperature difference matrices for these two animals are provided in~Fig.\ref{fig:matrices_outliers} in the Appendix.

\section{Conclusions}
We observed that characteristic thermal patterns of both horses and donkeys are usually associated with groups of ROIs (GORs) rather than an individual ROI. Based on this observation we defined a thermal pattern as a statistically significant difference between designated GORs for a given animal species. We have verified this significance both globally, for all data and locally, for individual animals.
We have shown how the majority of proposed thermal patterns are similar for both species. Noteworthy, the thermal patterns for donkeys are more uniform then for horses, and donkeys are individually more consistent with the global trend. Note that, the proposed thermal patterns compare data from one species or individual animals -- in general, for animals form our data set, average surface temperatures for horses are higher than for donkeys which may be related to differences in thermal properties of the skin and hair coat.

\section{Acknowledgments}
This work was conducted in the Veterinary Research Centre WULS (WCB) and the Center for Biomedical Research (CBB) supported by EFRR RPO WM 2007–2013. Authors are grateful to the Mrs. and Mr. Słupski, owners of the Mariaż agritourism farm in Lubochnia for letting them observe donkeys and horses and for their help on site.

\section{Appendix}

\begin{figure}[h!]
	\centering
	\includegraphics[width=0.24\linewidth]{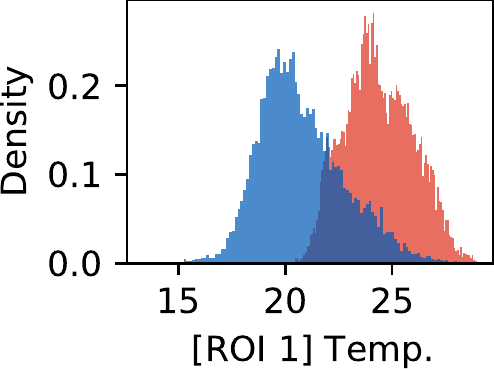}
	\includegraphics[width=0.24\linewidth]{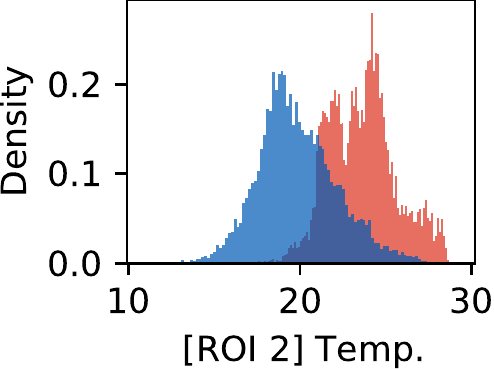}
	\includegraphics[width=0.24\linewidth]{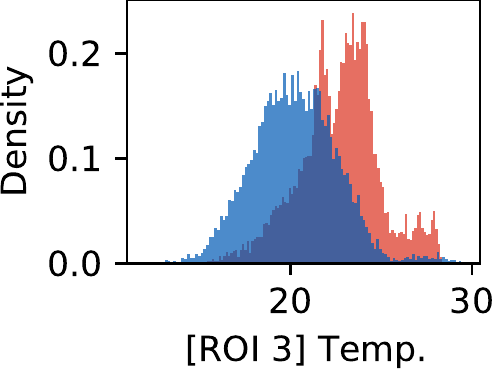}
	\includegraphics[width=0.24\linewidth]{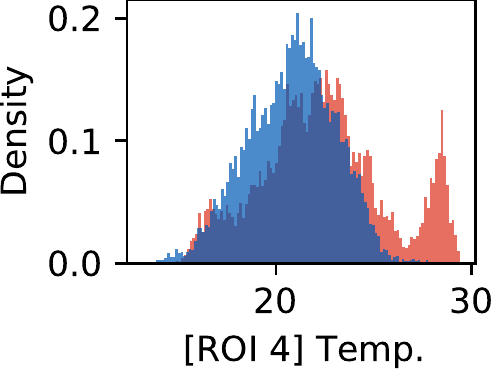}
	\includegraphics[width=0.24\linewidth]{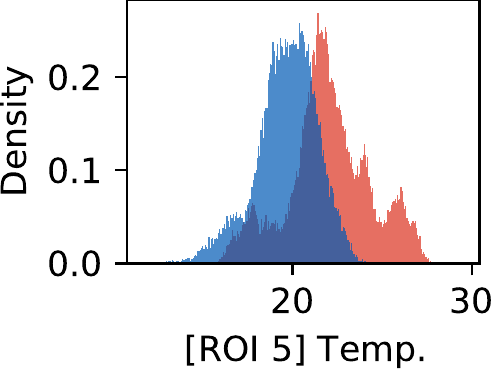}
	\includegraphics[width=0.24\linewidth]{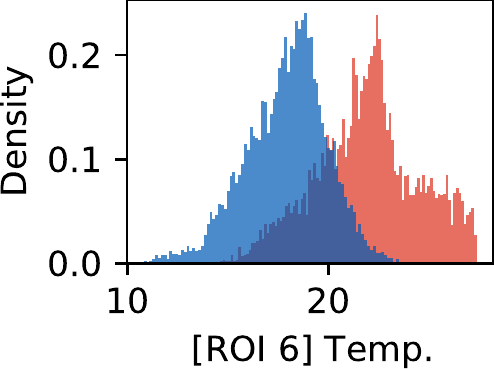}
	\includegraphics[width=0.24\linewidth]{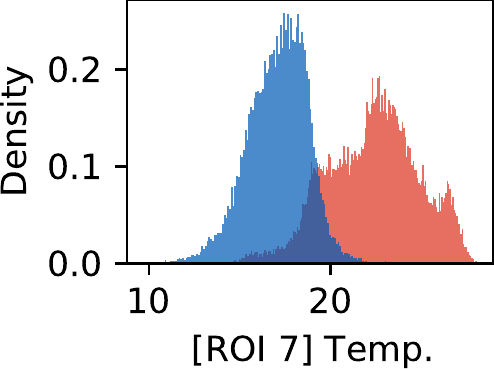}
	\includegraphics[width=0.24\linewidth]{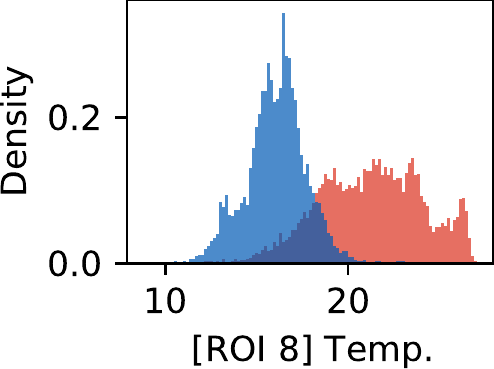}
	\includegraphics[width=0.24\linewidth]{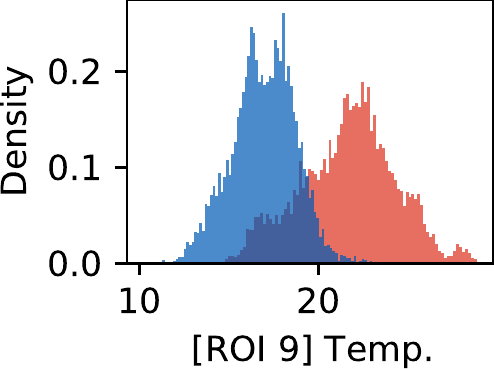}
	\includegraphics[width=0.24\linewidth]{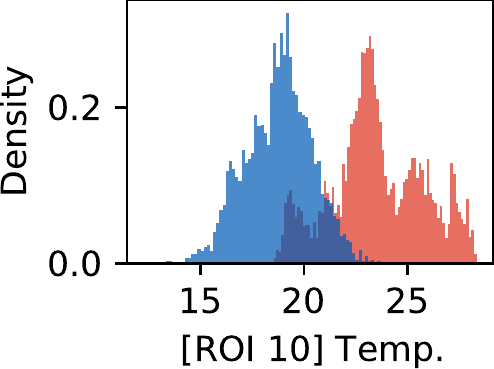}
	\includegraphics[width=0.24\linewidth]{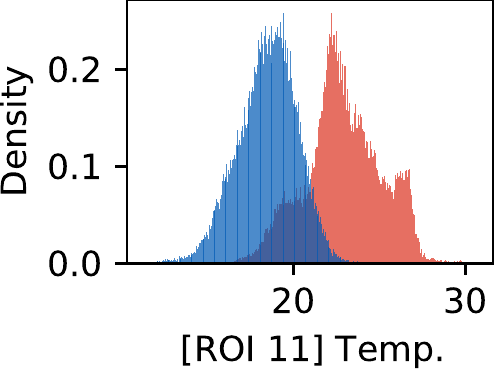}
	\includegraphics[width=0.24\linewidth]{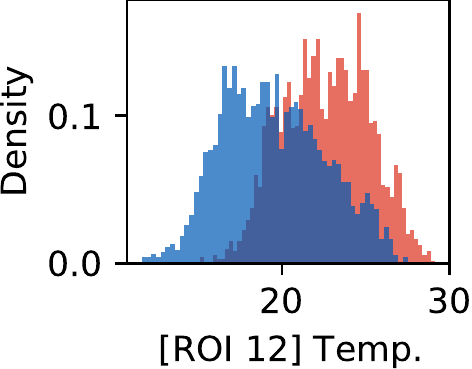}
	\includegraphics[width=0.24\linewidth]{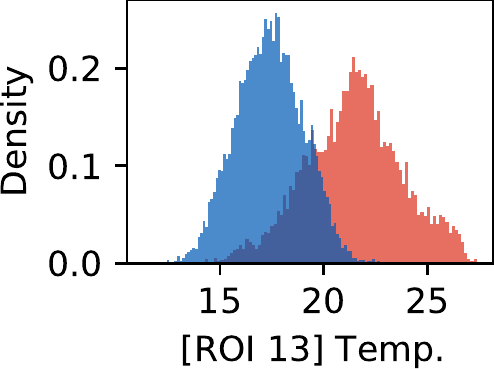}
	\includegraphics[width=0.24\linewidth]{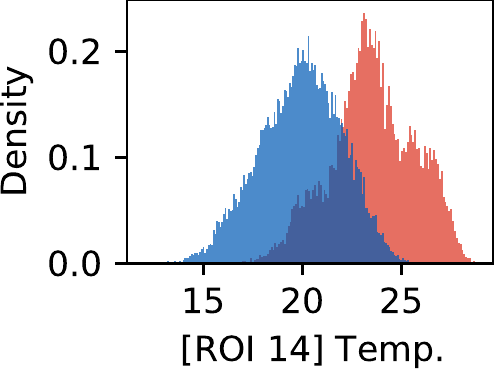}
	\includegraphics[width=0.24\linewidth]{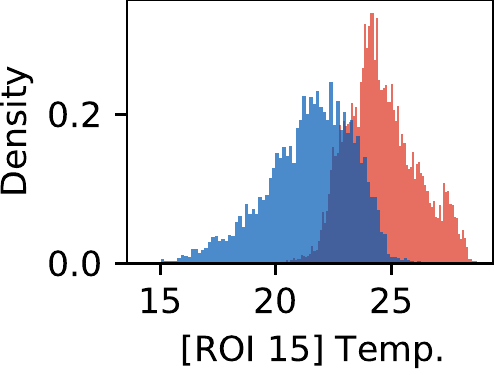}
	\includegraphics[width=0.24\linewidth]{images/histo/histo_15_small}
	\caption{Histograms of temperatures for all ROIs, the red colour denotes horses, the blue colour denotes donkeys. The last plot (the LR corner) presents the combined histogram for all ROIs.}
	\label{fig:histo_all}
\end{figure}

\begin{figure}[h!]
	\centering
	\begin{subfigure}[b]{0.49\textwidth}
    \includegraphics[width=1.0\linewidth]{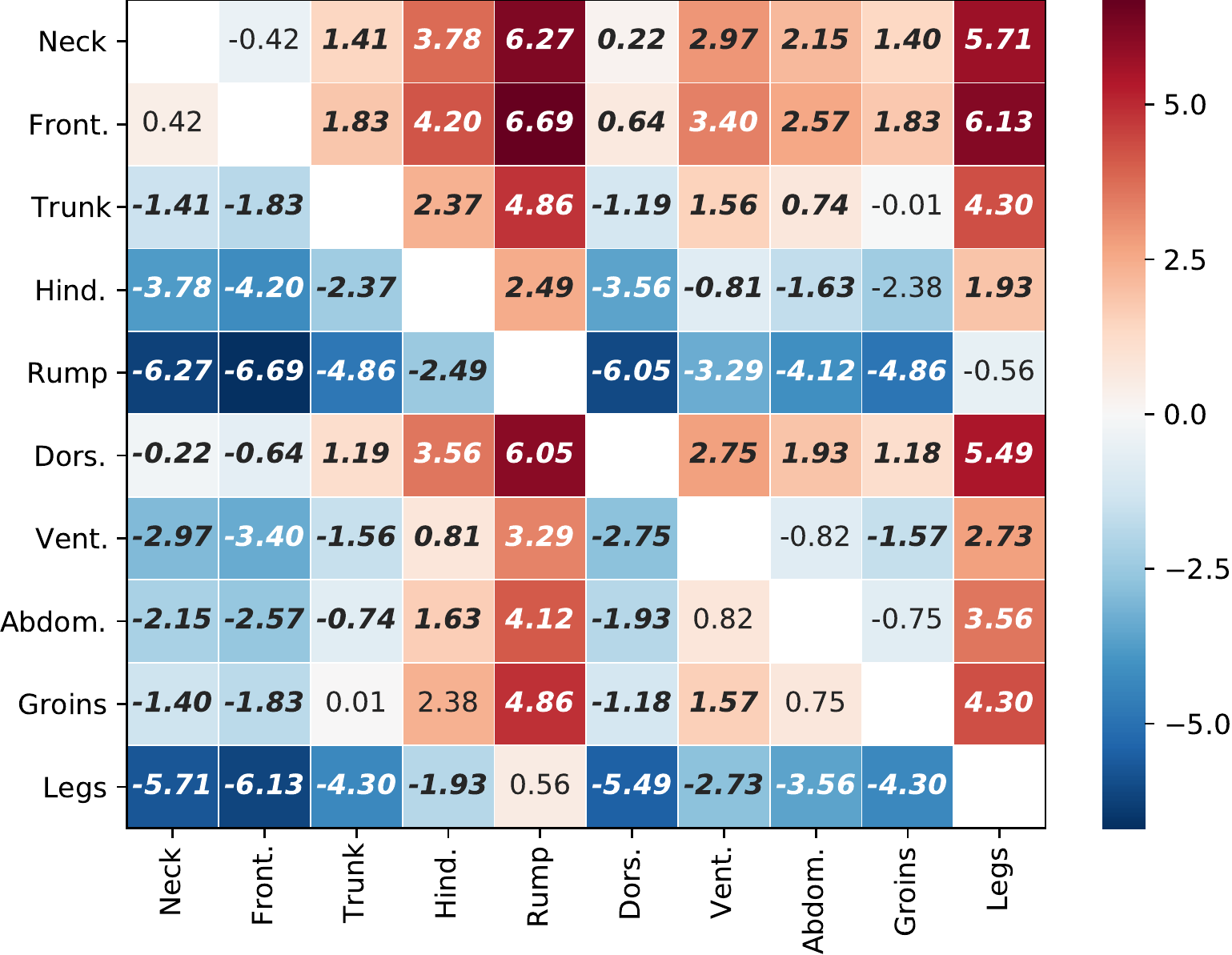}
    \caption{$\vect{M}^\Delta$, \emph{D.17}}
    \end{subfigure}
    \begin{subfigure}[b]{0.49\textwidth}
    \includegraphics[width=1.0\linewidth]{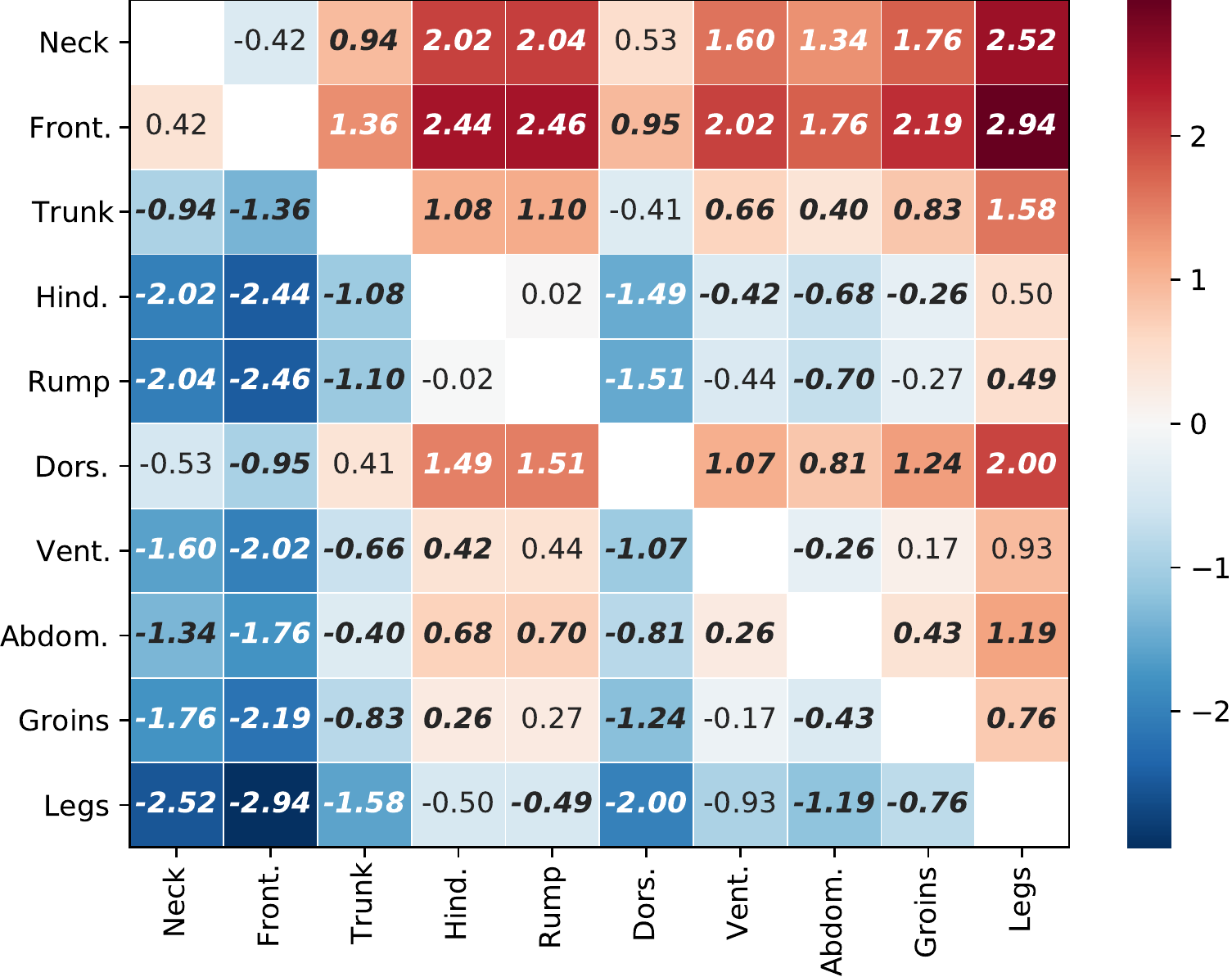}
    \caption{$\vect{M}^\Delta$, \emph{D.18}}
    \end{subfigure}
	\caption{Thermal patterns i.e. differences between GORs for the two donkeys identified as outliers (see. Sec.~\ref{sec:animals}): (a) Donkey \emph{D.17}; (b) Donkey \emph{D.18}. Bold font indicates the statistical significance of the difference for the given pattern.}
	\label{fig:matrices_outliers}
\end{figure}

\bibliographystyle{plain}
\bibliography{horses_vs_donkeys}

\begin{thebibliography}{10}

\bibitem{arruda2011thermographic}
TZ~Arruda, KE~Brass, and D~Flavio.
\newblock Thermographic assessment of saddles used on jumping horses.
\newblock {\em Journal of Equine Veterinary Science}, 31(11):625--629, 2011.

\bibitem{bachi2018changes}
B~B{\"a}chi, T~Wiestner, A~Stoll, NM~Waldern, I~Imboden, and MA~Weishaupt.
\newblock Changes of ground reaction force and timing variables in the course
  of habituation of horses to the treadmill.
\newblock {\em Journal of Equine Veterinary Science}, 63:13--23, 2018.

\bibitem{bartolome2013using}
E~Bartolom{\'e}, MJ~S{\'a}nchez, A~Molina, AL~Schaefer, I~Cervantes, and
  M~Valera.
\newblock Using eye temperature and heart rate for stress assessment in young
  horses competing in jumping competitions and its possible influence on sport
  performance.
\newblock {\em Animal: an International Journal of Animal Bioscience},
  7(12):2044, 2013.

\bibitem{becker2013cortisol}
M~Becker-Birck, A~Schmidt, M~Wulf, J~Aurich, A~Von~der Wense, E~M{\"o}stl,
  R~Berz, and C~Aurich.
\newblock Cortisol release, heart rate and heart rate variability, and
  superficial body temperature, in horses lunged either with hyperflexion of
  the neck or with an extended head and neck position.
\newblock {\em Journal of Animal Physiology and Animal Nutrition},
  97(2):322--330, 2013.

\bibitem{cetinkaya2012thermography}
MA~Cetinkaya and A~Demirutku.
\newblock Thermography in the assessment of equine lameness.
\newblock {\em Turkish Journal of Veterinary and Animal Sciences},
  36(1):43--48, 2012.

\bibitem{ciutacu2006igital}
O~Ciutacu, A~Tanase, and I~Miclaus.
\newblock Igital infrared thermography in assessing soft tissues injuries on
  sport equines.
\newblock {\em Bulletin of University of Agricultural Sciences and Veterinary
  Medicine Cluj-Napoca. Veterinary Medicine}, 63(1-2), 2006.

\bibitem{derrick2017comparing}
B~Derrick and P~White.
\newblock Comparing two samples from an individual likert question.
\newblock {\em International Journal of Mathematics and Statistics}, 18(3),
  2017.

\bibitem{eddy2001role}
AL~Eddy, LM~Van~Hoogmoed, and JR~Snyder.
\newblock The role of thermography in the management of equine lameness.
\newblock {\em The veterinary journal}, 162(3):172--181, 2001.

\bibitem{fonseca2006thermography}
BPA Fonseca, ALG Alves, JLM Nicoletti, A~Thomassian, CA~Hussni, and S~Mikail.
\newblock Thermography and ultrasonography in back pain diagnosis of equine
  athletes.
\newblock {\em Journal of Equine Veterinary Science}, 26(11):507--516, 2006.

\bibitem{alvarez2009back}
CB~G{\'o}mez~{\'A}lvarez, M~Rhodin, A~Bystr{\"o}m, W~Back, and PR~Van~Weeren.
\newblock Back kinematics of healthy trotting horses during treadmill versus
  over ground locomotion.
\newblock {\em Equine veterinary journal}, 41(3):297--300, 2009.

\bibitem{hinchcliff2013equine}
KW~Hinchcliff, AJ~Kaneps, and RJ~Geor.
\newblock {\em Equine Sports Medicine and Surgery E-Book}.
\newblock Elsevier Health Sciences, 2013.

\bibitem{hotelling1933analysis}
H~Hotelling.
\newblock Analysis of a complex of statistical variables into principal
  components.
\newblock {\em Journal of educational psychology}, 24(6):417, 1933.

\bibitem{kastberger2003infrared}
G~Kastberger and R~Stachl.
\newblock Infrared imaging technology and biological applications.
\newblock {\em Behavior Research Methods, Instruments, \& Computers},
  35(3):429--439, 2003.

\bibitem{machado2013standardization}
LFS Machado, RL~Dittrich, M~Pavelski, AMC da~F Afonso, I~Deconto, PT~Dornbusch,
  et~al.
\newblock Standardization of thermographic examination in joints of horses in
  training.
\newblock {\em Archives of Veterinary Science}, 18(4):40--45, 2013.

\bibitem{masko2019pattern}
M~Masko, L~Zdrojkowski, M~Domino, T~Jasinski, and Z~Gajewski.
\newblock The pattern of superficial body temperatures in leisure horses lunged
  with commonly used lunging aids.
\newblock {\em Animals}, 9(12):1095, 2019.

\bibitem{osthaus2018hair}
B~Osthaus, L~Proops, S~Long, N~Bell, K~Hayday, and F~Burden.
\newblock Hair coat properties of donkeys, mules and horses in a temperate
  climate.
\newblock {\em Equine veterinary journal}, 50(3):339--342, 2018.

\bibitem{pavelski2015infrared}
M~Pavelski, MS~Basten, E~Busato, and PT~Dornbusch.
\newblock Infrared thermography evaluation from the back region of healthy
  horses in controlled temperature room.
\newblock {\em Ci{\^e}ncia Rural}, 45(7):1274--1279, 2015.

\bibitem{purohit2009standards}
R~Purohit.
\newblock Standards for thermal imaging in veterinary medicine, 2009.

\bibitem{quaresma2013relationship}
M~Quaresma, R~Payan-Carreira, and SR~Silva.
\newblock Relationship between ultrasound measurements of body fat reserves and
  body condition score in female donkeys.
\newblock {\em The Veterinary Journal}, 197(2):329--334, 2013.

\bibitem{redaelli2019use}
V~Redaelli, F~Luzi, S~Mazzola, GD~Bariffi, M~Zappaterra, L~Nanni~Costa, and
  B~Padalino.
\newblock The use of infrared thermography (irt) as stress indicator in horses
  trained for endurance: A pilot study.
\newblock {\em Animals}, 9(3):84, 2019.

\bibitem{rosenmeier2012evaluation}
JG~Rosenmeier, AB~Strathe, and PH~Andersen.
\newblock Evaluation of coronary band temperatures in healthy horses.
\newblock {\em American journal of veterinary research}, 73(5):719--723, 2012.

\bibitem{satchell2015effects}
G~Satchell, M~McGrath, J~Dixon, T~Pfau, and R~Weller.
\newblock Effects of time of day, ambient temperature and relative humidity on
  the repeatability of infrared thermographic imaging in horses.
\newblock {\em Equine Veterinary Journal}, 47:13--14, 2015.

\bibitem{seeherman1991use}
HJ~Seeherman.
\newblock The use of high-speed treadmills for lameness and hoof balance
  evaluations in the horse.
\newblock {\em Veterinary Clinics of North America: Equine Practice},
  7(2):271--309, 1991.

\bibitem{silva2016relationships}
SR~Silva, R~Payan-Carreira, M~Quaresma, CM~Guedes, and AS~Santos.
\newblock Relationships between body condition score and ultrasound
  skin-associated subcutaneous fat depth in equids.
\newblock {\em Acta Veterinaria Scandinavica}, 58(1):37--42, 2016.

\bibitem{simon2006influence}
EL~Simon, EM~Gaughan, T~Epp, and M~Spire.
\newblock Influence of exercise on thermographically determined surface
  temperatures of thoracic and pelvic limbs in horses.
\newblock {\em Journal of the American Veterinary Medical Association},
  229(12):1940--1944, 2006.

\bibitem{smith1964applications}
WM~Smith.
\newblock Applications of thermography in veterinary medicine.
\newblock {\em Annals of the New York Academy of Sciences}, 121:248, 1964.

\bibitem{soroko2018infrared}
M~Soroko and K~Howell.
\newblock Infrared thermography: current applications in equine medicine.
\newblock {\em Journal of Equine Veterinary Science}, 60:90--96, 2018.

\bibitem{soroko2017effect}
M~Soroko, K~Howell, and K~Dudek.
\newblock The effect of ambient temperature on infrared thermographic images of
  joints in the distal forelimbs of healthy racehorses.
\newblock {\em Journal of Thermal Biology}, 66:63--67, 2017.

\bibitem{soroko2012use}
M~Soroko, E~Jodkowska, and M~Zab{\l}ocka.
\newblock The use of thermography to evaluate back musculoskeletal responses of
  young racehorses to training.
\newblock {\em Thermology International}, 22(3):114, 2012.

\bibitem{soroko2019evaluation}
M~Soroko, D~Zaborski, K~Dudek, K~Yarnell, W~G{\'o}rniak, and R~Vardasca.
\newblock Evaluation of thermal pattern distributions in racehorse saddles
  using infrared thermography.
\newblock {\em PloS one}, 14(8):e0221622, 2019.

\bibitem{tunley2004reliability}
BV~Tunley and FMD Henson.
\newblock Reliability and repeatability of thermographic examination and the
  normal thermographic image of the thoracolumbar region in the horse.
\newblock {\em Equine veterinary journal}, 36(4):306--312, 2004.

\bibitem{turner2001diagnostic}
TA~Turner.
\newblock Diagnostic thermography.
\newblock {\em Veterinary Clinics of North America: Equine Practice},
  17(1):95--114, 2001.

\bibitem{valera2012changes}
Mercedes Valera, Ester Bartolom{\'e}, Maria~Jos{\'e} S{\'a}nchez, Antonio
  Molina, Nigel Cook, and AL~Schaefer.
\newblock Changes in eye temperature and stress assessment in horses during
  show jumping competitions.
\newblock {\em Journal of Equine Veterinary Science}, 32(12):827--830, 2012.

\bibitem{maaten2008visualizing}
L~Van~der Maaten and G~Hinton.
\newblock Visualizing data using t-sne.
\newblock {\em Journal of machine learning research}, 9(Nov):2579--2605, 2008.

\bibitem{westermann2013effects}
S~Westermann, HHF Buchner, JP~Schramel, A~Tichy, and C~Stanek.
\newblock Effects of infrared camera angle and distance on measurement and
  reproducibility of thermographically determined temperatures of the
  distolateral aspects of the forelimbs in horses.
\newblock {\em Journal of the American Veterinary Medical Association},
  242(3):388--395, 2013.

\bibitem{zakari2018daily}
FO~Zakari, JO~Ayo, PI~Rekwot, MU~Kawu, and NS~Minka.
\newblock Daily rhythms of rectal and body surface temperatures in donkeys
  during the cold-dry (harmattan) and hot-dry seasons in a tropical savannah.
\newblock {\em International journal of biometeorology}, 62(12):2231--2243,
  2018.

\end{thebibliography}

\end{document}